\documentclass[fleqn,usenatbib]{mnras}
\usepackage{newtxtext,newtxmath}
\usepackage[T1]{fontenc}
\usepackage{ae,aecompl}
\usepackage{graphicx}   
\usepackage{amsmath}    
\usepackage{bbm}
\usepackage{amssymb}    

\title[Optimal tiling algorithm]{An optimised tiling pattern for multi-object spectroscopic surveys: application to the 4MOST survey}

\author[E. Tempel et al.]{
E. Tempel,$^{1}$\thanks{E-mail: elmo.tempel@ut.ee}
 T. Tuvikene,$^{1}$
 M.~M. Muru,$^{1}$
 R.~S. Stoica,$^{2}$
 T. Bensby,$^{3}$\newauthor
 C. Chiappini,$^{4}$
 N. Christlieb,$^{5}$
 M.-R.~L. Cioni,$^{4}$
 J. Comparat,$^{6}$
 S. Feltzing,$^{3}$\newauthor
 I. Hook,$^7$
 A. Koch,$^{8}$
 G. Kordopatis,$^{9}$
 M. Krumpe,$^{4}$
 J. Loveday,$^{10}$
 I. Minchev,$^{4}$\newauthor
 P. Norberg,$^{11}$
 B.~F. Roukema,$^{12,13}$
 J.~G. Sorce,$^{4,13}$
 J. Storm,$^{4}$
 E. Swann,$^{14}$\newauthor
 E.~N. Taylor,$^{15}$
 G. Traven,$^{3}$
 C.~J. Walcher,$^{4}$
 and R.~S. de~Jong$^{4}$
\vspace{1mm}
\\
$^{1}$Tartu Observatory, University of Tartu, Observatooriumi~1, 61602 T\~oravere, Estonia\\
$^{2}$Universit\'e de Lorraine, CNRS, IECL, 54000 Nancy, France\\
$^{3}$Lund Observatory, Department of Astronomy and Theoretical Physics, Box~43, SE-221~00 Lund, Sweden\\
$^{4}$Leibniz-Institut f\"ur Astrophysik Potsdam (AIP), An der Sternwarte 16, D-14482 Potsdam, Germany\\
$^{5}$Zentrum f\"ur Astronomie der Universit\"at Heidelberg, Landessternwarte, K\"onigstuhl 12, 69117 Heidelberg, Germany\\
$^{6}$Max-Planck-Institut f\"ur Extraterrestrische Physik (MPE), Giessenbachstr., D-85748 Garching, Germany\\
$^{7}$Physics Department, Lancaster University, Lancaster LA1~4YB, UK\\
$^{8}$Zentrum f\"ur Astronomie der Universit\"at Heidelberg, Astronomisches Rechen-Institut, M\"onchhofstr. 12, 69120 Heidelberg, Germany\\
$^{9}$Universit\'e C\^ote d'Azur, Observatoire de la C\^ote d'Azur, CNRS, Laboratoire Lagrange, France\\
$^{10}$Astronomy Centre, University of Sussex, Falmer, Brighton BN1 9QH, UK\\
$^{11}$Institute for Computational Cosmology and Centre for Extragalactic Astronomy, Department of Physics, Durham University,\\ South Road, Durham DH1 3LE, UK\\
$^{12}$Institute of Astronomy, Faculty of Physics, Astronomy and Informatics, Nicolaus Copernicus University, Grudziadzka 5,\\ 87-100 Toru\'n, Poland\\
$^{13}$Univ Lyon, Ens de Lyon, Univ Lyon1, CNRS, Centre de Recherche Astrophysique de Lyon UMR5574, F--69007, Lyon, France\\
$^{14}$Institute of Cosmology and Gravitation, University of Portsmouth, Burnaby Road, Portsmouth PO1 3FX, UK\\
$^{15}$Centre for Astrophysics and Supercomputing, Swinburne University of Technology, Hawthorn, VIC 3122, Australia
}

\pubyear{2020}

\begin{document}
\label{firstpage}
\pagerange{\pageref{firstpage}--\pageref{lastpage}}
\maketitle

\begin{abstract}
Large multi-object spectroscopic surveys require automated algorithms to optimise their observing strategy. One of the most ambitious upcoming spectroscopic surveys is the 4MOST survey. The 4MOST survey facility is a fibre-fed spectroscopic instrument on the VISTA telescope with a large enough field of view to survey a large fraction of the southern sky within a few years. Several Galactic and extragalactic surveys will be carried out simultaneously, so the combined target density will strongly vary. In this paper, we describe a new tiling algorithm that can naturally deal with the large target density variations on the sky and which automatically handles the different exposure times of targets. The tiling pattern is modelled as a marked point process, which is characterised by a probability density that integrates the requirements imposed by the 4MOST survey. The optimal tilling pattern with respect to the defined model is estimated by the tiles configuration that maximises the proposed probability density. In order to achieve this maximisation a simulated annealing algorithm is implemented. The algorithm automatically finds an optimal tiling pattern and assigns a tentative sky brightness condition and exposure time for each tile, while minimising the total execution time that is needed to observe the list of targets in the combined input catalogue of all surveys. Hence, the algorithm maximises the long-term observing efficiency and provides an optimal tiling solution for the survey. While designed for the 4MOST survey, the algorithm is flexible and can with simple modifications be applied to any other multi-object spectroscopic survey.
\end{abstract}



\section{Introduction}
\label{sec:intro}

\begin{figure*}
\centering
\includegraphics[width=\textwidth]{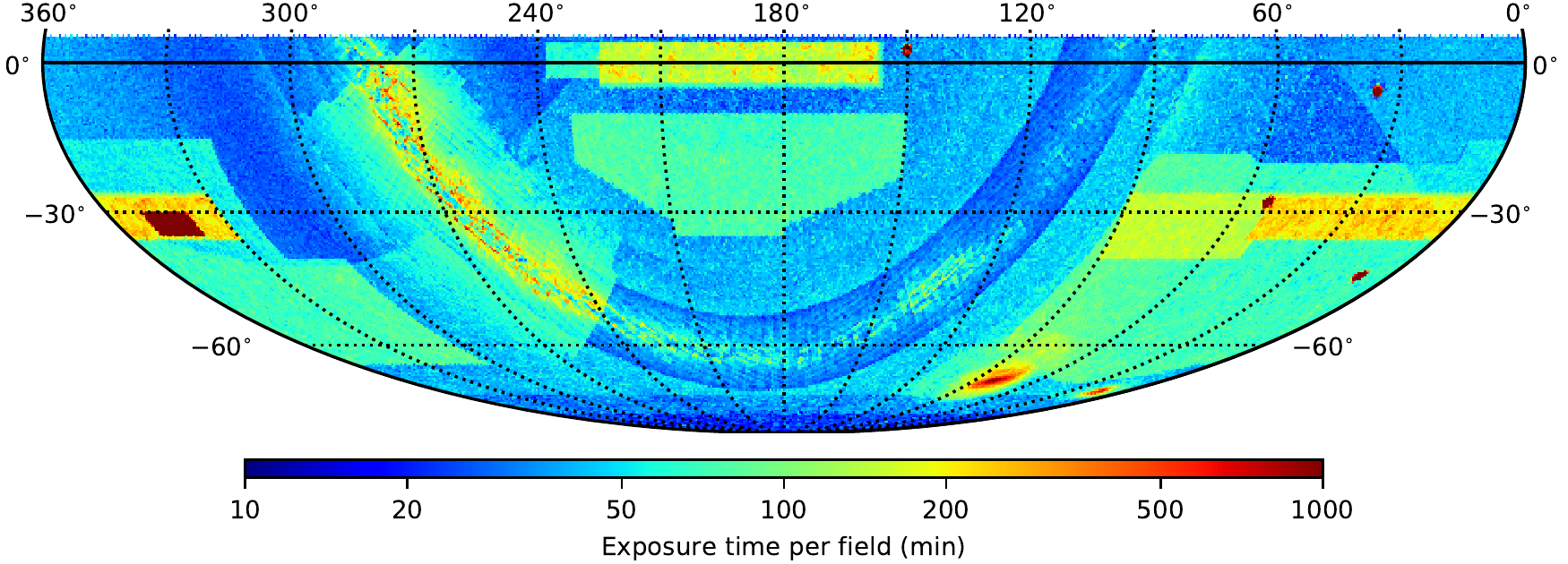}
        \caption{Required exposure-time map in equatorial coordinates for low-resolution (LR) targets, based on current 4MOST mock catalogues, both Galactic and extragalactic. The required exposure times and associated target densities vary significantly from one sky region to another. Exposure times have been calculated using the 4MOST Exposure Time Calculator assuming a fixed median seeing of 0.8 arcsec and airmass 1.2 for all targets. The targets are limited to declination between $-85$ and $+5$ degrees. The footprints of the different sub-surveys in 4MOST are clearly visible. The same set of targets is used in the examples presented in Section~\ref{sec:application}.}
        \label{fig:texp_map}
\end{figure*}

An integral part of the preparation of any multi-object spectroscopic survey is the construction of the tiling pattern (the set of centres and orientations on the sky of the observational field, ``tiles'') -- we need to know where to point the telescope and for how long each tile should be observed. In general, there are two approaches for finding an optimal tiling pattern. In the first approach, the tiling pattern is constructed on the fly, and the job of the tiling algorithm is to find the next telescope pointing, while taking into account already observed fields and targets. An example is the heuristic Greedy algorithm \citep{2010PASA...27...76R} that is used in the Galaxy And Mass Assembly (GAMA) survey \citep{2009A&G....50e..12D,2015MNRAS.452.2087L} and will be used in the Taipan survey \citep{2017PASA...34...47D}.

In the second approach, the tiling pattern is constructed before the survey starts, and is usually used to cover a given sky area uniformly. This approach is successfully used in the Two Degree Field Galaxy Redshift Survey \citep[2dFGRS,][]{2001MNRAS.328.1039C}, the Sloan Digital Sky Survey \citep[SDSS,][]{2003AJ....125.2276B}, the Six-degree Field (6dF) Galaxy Survey \citep{2004MNRAS.355..747J} and the WiggleZ survey \citep{2010MNRAS.401.1429D}. For these surveys, an adaptive tiling algorithm is used, where the uniform distribution of field centres is successively altered to more closely follow the target distribution. This algorithm is effective in providing uniform targeting completeness over the sky.

The Greedy algorithm \citep{2010PASA...27...76R} works very well for dense surveys, where the same sky region is visited several times. In contrast, an adaptive tiling algorithm is used when a given sky region needs to be covered with a minimum number of fields. In the 4MOST survey \citep{2019Msngr.175....3D,2019Msngr.175...12W}, both of these aspects must be optimized, so a new algorithm needs to be developed.

The 4MOST survey is a spectroscopic survey that will observe millions of targets covering almost the entire southern sky. The 4MOST survey consists of many sub-surveys covering different areas in the sky, which have very different number densities of targets. Fig.~\ref{fig:texp_map} shows the estimated exposure time in the sky based on the current 4MOST mock catalogues and the present survey strategy \citep{2019Msngr.175...17G}. In Fig.~\ref{fig:texp_map} we have combined the targets from the ten 4MOST consortium surveys: the Milky Way Halo Low-Resolution Survey \citep{2019Msngr.175...23H}, the Milky Way Halo High-Resolution Survey \citep{2019Msngr.175...26C}, the Milky Way Disc and Bulge Low-Resolution Survey \citep[4MIDABLE-LR,][]{2019Msngr.175...30C}, the Milky Way Disc and Bulge High-Resolution Survey \citep[4MIDABLE-HR,][]{2019Msngr.175...35B}, the eROSITA Galaxy Cluster Redshift Survey \citep{2019Msngr.175...39F}, the Active Galactic Nuclei Survey \citep{2019Msngr.175...42M}, the Wide-Area VISTA Extragalactic Survey \citep[WAVES,][]{2019Msngr.175...46D}, the Cosmology Redshift Survey \citep[CRS,][]{2019Msngr.175...50R}, the One Thousand and One Magellanic Fields Survey \citep[1001MC,][]{2019Msngr.175...54C}, and the Time-Domain Extra-galactic Survey \citep[TiDES,][]{2019Msngr.175...58S}.
The mock catalogues are based either on Gaia catalogues \citep{2016A&A...595A...1G,2018A&A...616A...1G} or on the Galaxia model of the Galaxy \citep{2011ApJ...730....3S}, or on MultiDark simulations augmented with models of galaxies and clusters \citep{2016MNRAS.457.4340K, 2019MNRAS.487.2005C}, or on TAO mocks \citep{2016ApJS..223....9B}, or on GALFORM mocks \citep{2000MNRAS.319..168C, 2012MNRAS.426.2142L}. They represent reasonably well each survey individually. Future work on mock catalogues should accurately reproduce the cross-correlation between surveys.
In the 4MOST surveys, most of the targets that will be observed are known from previous surveys and selected beforehand. The only exception is a small fraction of transients from the TiDES survey that will be selected based on live LSST observations. Since the number of transients is small and their spatial distribution is not clustered, we will ignore these targets in the current paper and we assume that all targets and their estimated exposure times are known.

\begin{figure}
\centering
\includegraphics[width=\columnwidth]{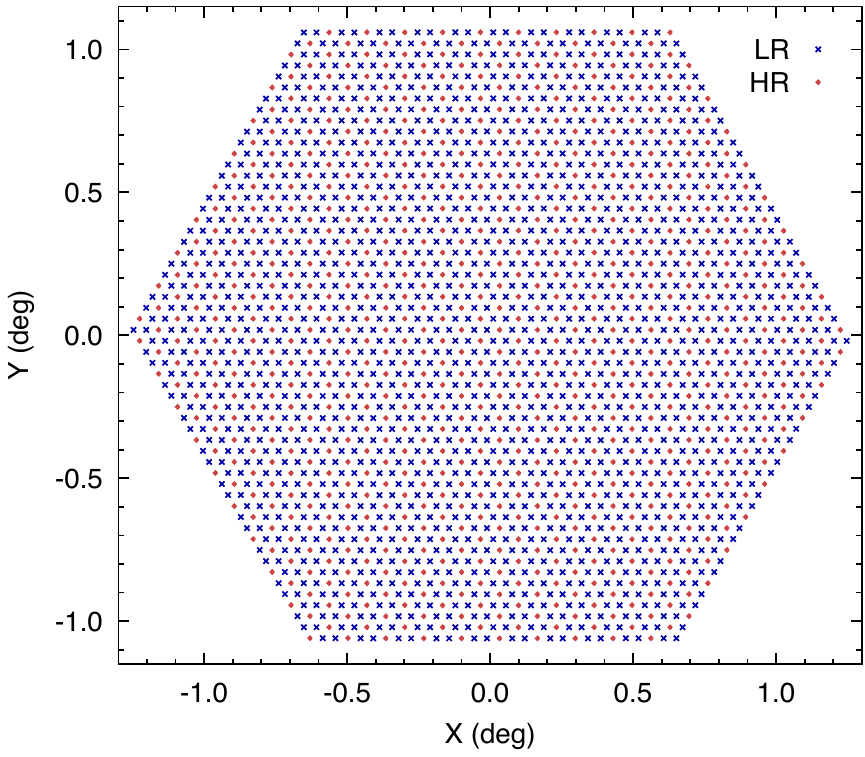}
        \caption{4MOST field of view with low-resolution (LR) and high-resolution (HR) spectrograph fibres at their home positions.}
        \label{fig:fov_fibres}
\end{figure}

The 4MOST field of view covers approximately four square degrees and is hexagonally shaped. It is covered by 1624 low-resolution (LR) and 812 high-resolution (HR) spectrograph fibres. On average, there are 391 LR and 196 HR spectrograph fibres per square degree. Fibres are placed with a regular pattern in the field of view and have some range of movement that allows them to be aligned to targets of interest (see Fig.~\ref{fig:fov_fibres}). \citet{Tempel:2019} gives a detailed overview of the capabilities and efficiency of the probabilistic fibre-to-target assignment algorithm developed specifically for the 4MOST survey. In order to apply the probabilistic targeting algorithm, we need a predefined tiling pattern that is optimised for the input targets and takes the constraints and requirements of the 4MOST facility and surveys into account. The most significant factor in determining the scientific impact of the 4MOST will be efficiency -- maximising fibre occupancy and minimising observational overheads.

The aim of this paper is to find an optimal tiling solution that increases survey efficiency. We propose an algorithm based on marked point processes. The idea is to model the tiling pattern as a marked point process, where each tile is considered as a free object, whose parameters (location, exposure time, etc.) need to be determined. A mathematically similar approach is successfully used to detect galaxy filaments \citep{2014MNRAS.438.3465T, 2016A&C....16...17T} and galaxy groups \citep{2018A&A...618A..81T} in spectroscopic galaxy surveys. Although the detection of cosmic web elements and finding the optimal tiling pattern are seemingly very different applications, mathematically both applications are pattern detection problems that can be tackled with the marked point process approach.

This paper is organised as follows. In Section~\ref{sec:problem} we describe the tiling challenge and define the inputs for, and the outputs of, the proposed tiling algorithm. In Section~\ref{sec:point_process} we describe the marked point process framework that we use to solve the optimal tiling problem. In Section~\ref{sec:application} we illustrate the proposed algorithm with examples. Conclusions are drawn in Section~\ref{sec:conclusions}.

\section{Tiling challenge}
\label{sec:problem}

The challenge we are facing in 4MOST survey preparation is how to most efficiently observe all required targets in the input catalogues, while maximising the fibre usage and minimising the total time (including overhead time) required to successfully observe the given set of targets. This can be considered as a tiling pattern optimisation problem. In the current paper we define a \textit{tile} as a single science exposure. In order to observe a given set of targets, we need a tiling pattern induced by the survey's input catalogues.

Each tile has a fixed sky coordinate (field centre) and instrument position angle. Several tiles with the same sky coordinates and position angle can be combined into one observing block (OB). Within one OB, tiles can have different exposure times and each tile has its own fibre-to-target configuration.

An algorithm that defines an optimal tiling pattern for the 4MOST survey should take into account the following aspects:
\begin{enumerate}
    \item Each tile has an individual exposure time that takes into account the requested exposure times of targets in the field of view. The exposure time is attached to each tile, assuming that it is observed in a fixed sky brightness condition (i.e., bright, grey or dark). This separates tiles into B/G/D groups\footnote{Separation into B/G/D (bright/grey/dark observing conditions) groups is somewhat arbitrary. In general, any number of groups can be used if it is necessary and if it helps to increase survey efficiency without over-complicating the problem.}.
    \item Tiles can be combined into OBs, which allows reducing the overhead time associated with telescope movement and field acquisition. One OB can contain one or many tiles with the same sky brightness condition (B/G/D) observed during one telescope pointing. The duration of one OB is limited by the total exposure time (approximately one hour per OB).
    \item In 4MOST there are two resolution modes -- high and low resolution. The fibre pattern for each of them is fixed and both of them are used simultaneously (see Fig.~\ref{fig:fov_fibres}). Each sub-survey specifies whether they want to use the high or the low resolution. The optimal tiling algorithm then aims to optimise both high- and low-resolution observations at the same time.
    \item Some sub-surveys can have specific requirements that affect the tiling pattern. For example, if a region in the sky is covered by several OBs (due to repeatability and/or high density of targets), then the centres of the OBs should preferentially avoid each other. Such a strategy will help to mitigate fixed fibre patterns. It will also tend to reduce visually striking contributions of the shape of the 4MOST field of view in the selection functions. Additionally, some sub-surveys require contiguous coverage of the sky, which translates to no gaps between tiles, while other sub-surveys wish to cover largest possible sky area and gaps between tiles are not a problem. An optimal tiling algorithm should be able to take into account these various scientific requirements from the different surveys.
\end{enumerate}

In general, to find the optimal tiling is a complicated problem that is interlinked with many other aspects of survey optimisation, including, for example, fibre-to-target assignment algorithm and the OB scheduling algorithm. The latter affects the division of tiles between different sky brightness conditions and the tentative exposure times of the tiles. For multiplex surveys such as the 4MOST survey, it is computationally unfeasible to solve all problems simultaneously. In the current paper, to reduce the complexity of the survey optimisation, the optimal tiling problem is solved independently with clearly defined inputs and outputs. The input data for the proposed tiling algorithm are described in Section~\ref{sec:tiling_input} and the output data are defined in Section~\ref{sec:tiling_output}.

The main aim of the proposed tiling algorithm is to find an optimal set of tiles that is required to observe a given set of targets. A probabilistic fibre to target assignment algorithm that uses the tiles as an input is described in \citet{Tempel:2019}. For the proposed algorithm, it is not important in which order the tiles are observed. The latter is a scheduling problem, which will be solved independently from the tiling and the fibre-to-target assignment algorithms.

\subsection{Input data for the tiling algorithm}
\label{sec:tiling_input}

The input data for the tiling algorithm are the following. We have a fixed set of targets, where for each target we have the following parameters:
\begin{itemize}
    \item \textbf{RA, Dec}: coordinates on the celestial sphere (``sky plane'').
    \item \textbf{Low or high resolution}: a flag that specifies whether the target should be observed with an LR or HR spectrograph fibre.
    \item \textbf{$T^\mathrm{B}_\mathrm{exp}$, $T^\mathrm{G}_\mathrm{exp}$, $T^\mathrm{D}_\mathrm{exp}$}: required exposure times\footnote{The required exposure time during real observations also depends on the sky transparency and seeing conditions. In this paper we ignore this effect and assume an average conditions everywhere.} of the target in bright, grey and dark sky conditions, respectively.
    \item \textbf{$f_\mathrm{compl}$}: the probability that the target should be successfully observed. To fulfil the survey science goals, some sub-surveys require only a fraction of targets from their input catalogues.
\end{itemize}

During the 4MOST five-year survey, approximately 32\% of the observing time is bright, 21\% is grey and 47\% is dark\footnote{The fraction of bright, grey and dark time depends on the the thresholds of sky brightness levels adopted. The current estimates are based on the ESO definitions for bright, grey and dark sky conditions (\url{https://www.eso.org/sci/observing/phase2_p101/ObsConditions.html}).}. The generated tiling pattern (total exposure time for D/G/B conditions) should roughly follow these fractions.

Because of the telescope and instrument design, the maximum exposure time for a single exposure is limited. In the current paper, we assume that the maximum exposure time is 30 minutes. Additionally, the total time (summed exposure times plus overheads) for a single OB is typically around one hour. In the current paper, we adopt a maximum OB length of 75 minutes, which is also feasible. This allows the observation of two 30~min science exposures in a single OB.

Because of the overhead time for each exposure/tile and for each OB\footnote{In the current paper, each OB has an overhead of 3.5 minutes and each exposure/tile has an additional overhead of 4.4 minutes (\url{https://www.4most.eu/cms/facility/capabilities/}). These overhead times are current estimates and might change before the 4MOST survey starts.}, several tiles/exposures are combined into one OB, which helps to reduce the total overhead time. Additionally, since overhead is constant (it does not depend on exposure times), this choice should reduce the total number of tiles, which tends to yield longer exposures.

\subsection{Expected output of the tiling algorithm}
\label{sec:tiling_output}

The tiling algorithm should find the tiling pattern that allows optimal observation of the given set of targets. The output of the tiling algorithm is the list of OBs, where for each OB we have the following parameters:
\begin{itemize}
    \item \textbf{RA, Dec}: the coordinates of the centre of an OB.
    \item \textbf{Position angle}: an angle determining the rotation of the field (hexagon) in the sky.
    \item \textbf{B/G/D flag}: a flag specifying whether the OB should be preferentially observed during bright, grey or dark sky conditions.
    \item \textbf{List of tiles and exposure times}: one or several tiles/exposures. The tiling algorithm should give the number of tiles for each OB. Tiles in one OB can have different exposure times. The algorithm should determine the expected exposure time for each tile. The sum of these exposures plus the overhead time is the total time for a single OB.
\end{itemize}

The distribution of OBs/tiles in the sky and exposure times per tiles should allow the observation of the required set of targets in the input catalogue, while minimising unused/wasted observational time (e.g., empty fibres, overexposure). In general, the optimal tiling solution allows to successfully observe the required set of targets with a minimum amount of time.

\subsection{Proposed tiling algorithm in a nutshell}

In the next Section, we describe the mathematical framework of the proposed optimal tiling algorithm and provide all the necessary details. To help understand the general concept of the algorithm, here we present a general outline of the process.

In the proposed algorithm, we model the tiling pattern as a marked point process (see Sections~\ref{sec:math_setup} and \ref{sec:model_construction}), where the number of tiles, together with the location and exposure time of each tile, are free parameters. An optimal tiling pattern is defined via an energy function: the global minimum of this energy function defines the optimal tiling pattern. We define the energy function as a sum of individual components, where each optimises a certain aspect of the tiling pattern. The most important energy function component is computed using a statistical fibre-to-target assignment algorithm. This allows us to compare the generated tiling pattern with the targets in the input catalogue, in order to estimate the time that would still be needed in order to observe the required targets in the input catalogue (``missing'' time) and to estimate the time that remains unused due to empty fibres. Additional energy function components are used to minimise the total overhead time and to divide the tiles between predefined sky conditions, while minimising the total time that is necessary in order to observe all required targets in the input catalogue. We also define an energy function with components that allow us to define the interactions between tiles in a way that potentially minimises the impact of the fixed tiling pattern on the final selection function of the survey.

The optimisation challenge we are facing involves a large number of parameters, whereas the number of free parameters (the number of tiles) is itself a free parameter. The proposed algorithm finds itself the number of tiles. The minimisation of the energy function is achieved via a simulated annealing algorithm, which is a global optimisation method that avoids local minima.

\section{Marked point process for determining optimal tiling pattern}
\label{sec:point_process}

\subsection{Mathematical set-up of the problem}
\label{sec:math_setup}

The key assumption of our proposed algorithm is that the tiling pattern is a configuration of random interacting objects driven by the probability density of a marked point process. The solution of the optimal tiling pattern is given by the construction and manipulation of such a probability density. The probability density we propose takes into account all the observational constraints and requirements from all surveys. Statistical inference using this probability density is done using Markov-chain Monte Carlo (MCMC) techniques. Such a probability density can be written as
\begin{equation}
        p(\mathbf{y}\,|\,\theta) \propto \exp\left[-U(\mathbf{y}\,|\,\theta)\right],
        \label{eq:gibbs}
\end{equation}
where $U(\mathbf{y}\,|\,\theta)$ is the energy function, $\mathbf{y}$ is the pattern of objects (tiles in the sky) and $\theta$ is the vector of model parameters. The marked point processes driven by probability densities in Eq.~\eqref{eq:gibbs} are known in the literature as a Gibbs point processes. The energy function $U(\mathbf{y}\,|\,\theta)$ can be further written as the sum of several components that take into account different aspects of the optimisation problem (see Section~\ref{sec:model_construction}).

The Bayesian framework allows the introduction of the knowledge regarding the parameters via a posterior distribution $p(\theta)$. This allows writing the joint distribution of the tiling pattern and the model parameters:
\begin{equation}
  p(\mathbf{y},\theta) = p(\mathbf{y}\,|\,\theta)\,p(\theta).
        \label{eq:probdens}
\end{equation}
A joint tiling pattern and parameter estimator is given by the maximum of the probability density~(\ref{eq:probdens}):
\begin{equation}
        (\hat{\mathbf{y}},\hat{\theta}) = \mathrm{arg}\,\max_{\Omega\times\Theta}\, p(\mathbf{y},\theta) =
        \mathrm{arg}\,\max_{\Omega\times\Theta}\, p(\mathbf{y}\,|\,\theta) \,p(\theta),
        \label{eq:estimator}
\end{equation}
where $\Omega$ is the pattern configuration space and $\Theta$ represents the parameter space. The estimator given by (\ref{eq:estimator}) can be computed using a simulated annealing algorithm \citep{vanLieshout:94,Stoica:05}.

For simplicity and in order to reduce the computational cost, most of the model parameters $\theta$ are fixed during the Monte Carlo simulation. In the current paper, the estimation of these parameters is done using an educated guess and via trial and error, whenever necessary. The free parameters of the model are described in Section~\ref{sec:model_construction} and the parameter values used in the current paper are given in Table~\ref{tab:parameters}. The model parameters $\theta$ can be estimated if the tiling pattern is available following \citet{Stoica:17} and the references therein.

\subsection{Model construction}
\label{sec:model_construction}

Let $W$ be a spatial observation window of Lebesgue measure $\nu(W)$. In the current paper, $W$ is a finite region in the sky plane (sky area reachable by the 4MOST facility). A simple point process on $W$ is a finite random configuration of points $x_i \in W$, $i=1,\dots,n$ such that $x_i\neq x_j$ whenever $i\neq j$, where $n$ is the number of points in a point process. Characteristics or marks can be attached to the points via a probability distribution. A finite random configuration of marked points is a marked point process if the distribution of only the locations is a simple point process. For further reading on marked point processes we recommend the monographs by \citet{vanLieshout:00} and \citet{MollWaag:04}. In the current paper, tiles are considered as marked points and they are modelled as a marked point process.

The generating object (a marked point) of the tiling pattern is given by a tile $y = (\alpha,\delta,\mathrm{PA},i_\mathrm{BGD},T_\mathrm{exp})$. The tile centre coordinates are given by $\alpha,\delta \in W$, where $\alpha,\delta$ are right ascension and declination in the sky. The mark is represented by the following parameters: $\mathrm{PA}$ is a position angle, $i_\mathrm{BGD}$ is the sky condition flag and $T_\mathrm{exp}$ gives the exposure time of the tile. To find the optimal tiling pattern means to find the set of tiles $\mathbf{y} = y_1,y_2,\ldots,y_{N_\mathrm{tiles}}$ that are needed to observe a given set of targets $\mathbf{t} = t_1,t_2,\ldots,t_{N_\mathrm{tar}}$. While the number of targets $N_\mathrm{tar}$ and parameters of targets (see Section~\ref{sec:tiling_input}) are known, the number of tiles $N_\mathrm{tiles}$ and parameters of each tile ($\alpha,\delta,\mathrm{PA},i_\mathrm{BGD},T_\mathrm{exp}$) are the subject of optimal tiling pattern determination described in Section~\ref{sec:problem}.

The optimal tiling estimator is defined by the tiling configuration that maximises the probability density in Eq.~\eqref{eq:gibbs} as it minimises the corresponding energy function $U(\mathbf{y}\,|\,\theta)$. For the problem at hand, the energy function $U(\mathbf{y}\,|\,\theta)$ is constructed as follows:
\begin{eqnarray}
    U(\mathbf{y}\,|\,\theta) = U_\mathrm{targets}(\mathbf{y}\,|\,\theta) + U_\mathrm{overhead}(\mathbf{y}\,|\,\theta) + \nonumber \\ U_\mathrm{tiles}(\mathbf{y}\,|\,\theta) + U_\mathrm{BGD}(\mathbf{y}\,|\,\theta),
    \label{eq:energy_function}
\end{eqnarray}
where each component in the energy function takes into account different aspects in the optimal tiling challenge. The energy function $U(\mathbf{y}\,|\,\theta)$ is calculated for a given set of tiles $\mathbf{y}$ and using a fixed set of targets $\mathbf{t}$. Each component of the energy function is described below in detail.
The function $U_\mathrm{targets}(\mathbf{y}\,|\,\theta)$ takes into account the exposure times of targets and is used to minimise the summed exposure time of tiles that is needed to observe a given set of targets $\mathbf{t}$; $U_\mathrm{overhead}(\mathbf{y}\,|\,\theta)$ minimises the overhead associated with each OB and individual exposures; $U_\mathrm{tiles}(\mathbf{y}\,|\,\theta)$ is used to optimise the placement of tiles with respect to each other; and $U_\mathrm{BGD}(\mathbf{y}\,|\,\theta)$ is introduced to comply with the available fraction of observational time in bright, grey or dark sky conditions.

The definition of the terms of each energy function component (see below) together with the values of the parameters lead to a locally stable model. This means that the contribution to the general energy function of a new tile to an existing configuration is bounded below. This property implies the integrability of the model. The local stability is also required in order to obtain the required convergence properties for the simulation algorithm of the model \citep{vanLieshout:00, vanLieshout:03, MollWaag:04}. In the following, we describe the implementation details of each of the energy function components and of the MCMC simulation method.

\subsubsection{$U_\mathrm{targets}(\mathbf{y}\,|\,\theta)$}
\label{sec:Utargets}

This is the most important component of the energy function. This component ensures that the targets in the input catalogue are observed efficiently. The energy function $U_\mathrm{targets}(\mathbf{y}\,|\,\theta)$ is based on all targets and all tiles in the sky. While minimizing this energy we find the best tiling that allows optimal observation of a given set of targets. This is defined as

\begin{equation}
    U_\mathrm{targets}(\mathbf{y}\,|\,\theta) =
    \frac{1}{A(\mathrm{FoV})} \iint\limits_{S} U^s_\mathrm{targets}(\mathbf{y}\,|\,\theta)\mathrm{d}s,
    \label{eq:Utargets_int}
\end{equation}
\begin{eqnarray}
    U^s_\mathrm{targets}(\mathbf{y}\,|\,\theta) &=&
        \left[c_\mathrm{miss}T^s_\mathrm{miss} + c_\mathrm{wasted}T^s_\mathrm{wasted}\right] \nonumber \\
        \quad \mathrm{for} &&
        \left\{t\in\mathbf{t}:\|t-s\|<s_\mathrm{max}\right\},
        \label{eq:Us_targets}
\end{eqnarray}
where $S\in W$ is the region in the sky where targets are located. The inverse of the normalization constant in front of the surface integral, $A(\mathrm{FoV})$, is the area of one 4MOST field of view. This gives the energy (missing and wasted time) as an average quantity per one field of view.
For a region $s$ in the sky, the function $U^s_\mathrm{targets}$ is estimated based on targets closer than $s_\mathrm{max} = 0.1$~deg from the centre of region $s$. In a circle with radius 0.1~deg there are on average 12 LR and 6 HR spectrograph fibres. For simplicity, the integral in Eq.~\eqref{eq:Utargets_int} is estimated as a sum over HEALPix\footnote{\url{https://healpix.jpl.nasa.gov}} pixels \citep{Healpix99} in the $(H=4,N=3)$ member of the HEALPix family of equal-area projections from the sky to the plane \citep{CR07}. We use the HEALPix Nside parameter of 1024, which gives around 1300 pixels in one 4MOST field of view. The $s_\mathrm{max}$ defines the smoothing scale for the $U^s_\mathrm{targets}$ energy function component.

In Eq.~\eqref{eq:Us_targets}, $T^s_\mathrm{miss}$ is the exposure time that is missing in order to observe all targets in region $s$ and $T^s_\mathrm{wasted}$ is the observational time that is not used for science targets. Wasted time counts the time that is not used at all for science targets (e.g., empty fibres) and counts the time over which science targets were over-exposed. Positive constants $c_\mathrm{miss}\geq 0$ and $c_\mathrm{wasted}\geq 0$ can be used to fine tune the balance between missing and wasted observations.

In the current paper, the missing and wasted time is estimated as:
\begin{equation}
    T^s_\mathrm{miss} = \sum\limits_{\mathrm{X}\in[\mathrm{LR,HR}]} c_\mathrm{X}\left[T^s_\mathrm{req,X} - T^s_\mathrm{obs,X}\right],
    \label{eq:tmissing}
\end{equation}
\begin{equation}
    T^s_\mathrm{wasted} = \sum\limits_{\mathrm{X}\in[\mathrm{LR,HR}]}  c_\mathrm{X}\left[T^s_\mathrm{over-exp,X} + T^s_\mathrm{not-used,X}\right],
    \label{eq:twasted}
\end{equation}
where $T^s_\mathrm{req}$ is the required exposure time in a region $s$ in order to observe all targets in this region; $T^s_\mathrm{obs}$ is the exposure time that was actually used to observe the targets in this region. The term $T^s_\mathrm{over-exp}$ takes into account the over-exposure of targets and $T^s_\mathrm{not-used}$ gives the time that was not used for science targets (time lost because of empty fibres). The missing and wasted time is calculated separately for low and high resolution. The parameters $c_\mathrm{LR}$ and $c_\mathrm{HR}$ can be used to control the relative importance of LR and HR targets and fibres. In the current paper, we set $c_\mathrm{LR}=2/3$ and $c_\mathrm{HR}=1/3$, so that they reflect the number density of LR and HR spectrograph fibres.

To calculate the quantities $T^s_\mathrm{req}$, $T^s_\mathrm{obs}$, $T^s_\mathrm{over-exp}$ and $T^s_\mathrm{not-used}$, we have to assign fibres to targets. \citet{Tempel:2019} describes a probabilistic fibre-to-target assignment algorithm. However, this algorithm is computationally expensive and in practice it cannot be used in the proposed tiling algorithm. To generate an optimal tiling pattern, we will use a simplified (and computationally faster) version of the fibre-to-target assignment algorithm. The simplified version described below does not assign real fibres to targets; it is only used to statistically mimic the fibre-to-target assignment. Hence, the simplified targeting used in the current paper does not replace the need for a probabilistic targeting such as the one proposed in \citet{Tempel:2019}. In the current paper, the calculation of $T^s_\mathrm{req}$, $T^s_\mathrm{obs}$, $T^s_\mathrm{over-exp}$ and $T^s_\mathrm{not-used}$ is performed as described below. The calculation of these quantities is the same for LR and HR targets, except that the fibre density is different for LR and HR fibres.

\paragraph*{Calculation of  $T^s_\mathrm{req}$.} To estimate the required exposure time in a region $s$, we assume that any fibre can be placed on any target in that region. The $T^s_\mathrm{req}$ for LR targets is estimated as
\begin{equation}
    T^s_\mathrm{req,LR} = \frac{ 1 }{N^s_\mathrm{LR,fib}} \sum\limits_{t\in \mathbf{t}_s^\mathrm{LR}} T_\mathrm{exp}^\mathrm{D}(t) \cdot f_\mathrm{compl}(t) \,,
    \label{eq:texp_req}
\end{equation}
where the summation is over LR targets $t$ that belong to region $s$, in the sense that they are closer than $s_\mathrm{max}$ to the centre of region $s$ in the sky. The set of LR targets that belong to region $s$ is designated as $\mathbf{t}_s^\mathrm{LR}$. The parameter $N^s_\mathrm{LR,fib}$ gives the average number of LR fibres in region $s$. It is estimated as
\begin{equation}
    N^s_\mathrm{LR,fib} = c_\mathrm{sci\_fib}\cdot \rho_\mathrm{fib}^{\mathrm{LR}} \cdot A(s) \,,
\end{equation}
where $\rho_\mathrm{fib}^{\mathrm{LR}}$ defines the average LR fibre density in one field of view, $A(s)$ gives the area of region $s$ and parameter $c_\mathrm{sci\_fib}\in [0\ldots 1]$ defines the fraction of fibres that are available for science targets\footnote{The fraction is less than one, because some fibres are used for calibration or are allocated as sky fibres.}.

The calculation of $T^s_\mathrm{req}$ for HR targets is the same, except that we use HR targets and HR fibre density $\rho_\mathrm{fib}^{\mathrm{HR}}$. In the 4MOST facility, the average LR fibre density is $\rho_\mathrm{fib}^{\mathrm{LR}}=391~\mathrm{sq\,deg}^{-1}$ and for HR it is $\rho_\mathrm{fib}^{\mathrm{HR}}=196~\mathrm{sq\,deg}^{-1}$.

In Eq.~\eqref{eq:texp_req}, we use exposure times for the dark sky condition. Hence, Eq.~\eqref{eq:texp_req} gives the minimum exposure time required to observe all targets in the dark sky condition, assuming perfect fibre-to-target assignment without any loss. Effectively, the perfect fibre-to-target assignment is only used to estimate the missing time, $T^s_\mathrm{miss}$. The simplified fibre-to-target assignment described below takes into account the sky conditions associated with each OB and different sky conditions are also included to calculate the $T^s_\mathrm{obs}$.

\paragraph*{Calculation of  $T^s_\mathrm{obs}$.} The observed exposure time $T^s_\mathrm{obs}$ counts the time that is used to observe targets in a region~$s$. It is estimated as
\begin{equation}
    T^s_\mathrm{obs,LR} = \frac{ 1 }{N^s_\mathrm{LR,fib}} \sum\limits_{t\in \mathbf{t}_s^\mathrm{LR}}  \min\left[1.0,f_\mathrm{obs}(t)\right] \cdot T_\mathrm{exp}^\mathrm{D}(t) \cdot f_\mathrm{compl}(t) \,,
    \label{eq:texp_obs}
\end{equation}
where notations are the same as in Eq.~\eqref{eq:texp_req} and $f_\mathrm{obs}(t)$ gives the completion fraction for target $t$. If the summed exposure time of a target $t$ is equal or larger than the requested exposure time, then effectively $f_\mathrm{obs}(t)=1.0$.

The completion fraction $f_\mathrm{obs}(t)$ for target $t$ is estimated using the simplified fibre-to-target assignment. For that we start with all tiles that cover the region $s$. The set of these tiles is designated $\mathbf{y}_s$. We assume that all fibres in region $s$ can be used for all targets in that region. This is an approximation, but since region $s$ is relatively small, it is a statistically unbiased approach. Since the exposure times of tiles $y\in\mathbf{y}_s$ can be different, the assignment of fibres/tiles to targets itself is an optimisation problem. In the current paper, we do not solve this extra optimisation problem, for reasons of minimising the computational time. We instead use a simple scheme that provides an efficient enough solution. The simplified fibre-to-target assignment for LR targets in region $s$ is done as follows.
\begin{enumerate}
    \item Initialize all tiles $y\in\mathbf{y}_s$ in the region $s$. All tiles have the same fixed number of LR fibres for science targets, $N_\mathrm{LR,fib}$. Initially, for all tiles $y\in\mathbf{y}_s$, $N_\mathrm{fib}^\mathrm{alloc}(y)=0$, which is the number of allocated fibres in tile $y$.
    \item Select all LR targets $t\in\mathbf{t}_s^\mathrm{LR}$ in the region $s$. Sort all targets $t\in\mathbf{t}_s^\mathrm{LR}$ for descending order of target exposure time $T_\mathrm{exp}^\mathrm{D}(t)$. For each target, set the completion fraction to zero, $f_\mathrm{obs}(t)=0$.
    \item For each target, set the overexposure fraction to zero, $f_\mathrm{over-exp}(t)=0$.
    \item Loop over targets $t$ in the region $s$, starting from the target with the longest exposure time $T_\mathrm{exp}^\mathrm{D}(t)$. For each target, allocate fibres as follows.
    \begin{enumerate}
        \item For target $t$, calculate completion fraction when observing with field $y\in\mathbf{y}_s$. Completion fraction for target $t$ in field $y$ is estimated as $f_y(t) = T^{y}_\mathrm{exp}/T^\mathrm{BGD}_\mathrm{exp}$, where $T^{y}_\mathrm{exp}$ is tile exposure time and for the target exposure time $T^\mathrm{BGD}_\mathrm{exp}$ we use the sky condition that matches with the tile condition $i^y_\mathrm{BGD}$.
        \item Look through all tiles that are available in the region $s$. Tile $y$ is available for target $t$, if $N_\mathrm{fib}^\mathrm{alloc}(y) < N_\mathrm{LR,fib}$ and target $t$ is not yet observed with tile $y$.
        \item If there exists a tile $y$ for which  $f_\mathrm{obs}(t) + f_y(t)>1.0$, then assign target $t$ to the tile $y$, where $f_\mathrm{obs}(t) + f_y(t)$ is the lowest. For that tile, increase the fibre allocation $N_\mathrm{fib}^\mathrm{alloc}(y)$ by $f_\mathrm{compl}$. For the target $t$, set $f_\mathrm{obs}(t) = 1.0$. For target $t$, set the over-exposure fraction $f_\mathrm{over-exp}(t)= f_\mathrm{obs}(t) + f_y(t) - 1.0$. Go to the next target in the region $s$.
        \item If the condition in (c) is not met then allocate target $t$ to a tile, where $f_y(t)$ is the largest. For that tile, increase the fibre allocation $N_\mathrm{fib}^\mathrm{alloc}(y)$ by $f_\mathrm{compl}$. For the target $t$, set $f_\mathrm{obs}(t) = f_\mathrm{obs}(t) + f_y(t)$.
        \item If for any tile ($y\in\mathbf{y}_s$) $N_\mathrm{fib}^\mathrm{alloc}(y)$ is larger than the number of available fibres $N_\mathrm{fib}$, remove this tile from the available tile list.
        \item Go to the point (b) and add target $t$ to another tile $y$.
        \item If there are no tiles available (for example when the requested target exposure time is larger than the total exposure time in this region), go to the next target in the region $s$.
    \end{enumerate}
\end{enumerate}
Since for each target we estimate the target completion fraction $f_\mathrm{obs}(t)$, Eq.~\eqref{eq:texp_obs} allows us to estimate the actual observed time for dark sky conditions, and combining this with Eq.~\eqref{eq:texp_req}, allows us to directly estimate the missing observational time $T^s_\mathrm{miss}$, see Eq.~\eqref{eq:tmissing}.

\begin{figure}
\centering
\includegraphics[width=\columnwidth]{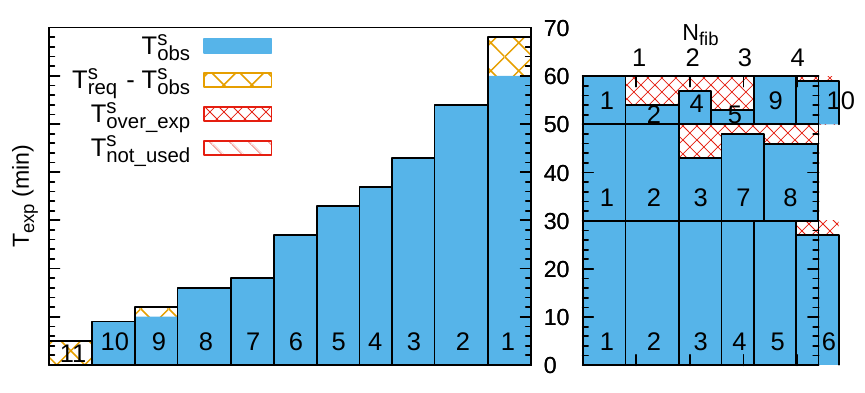}
\includegraphics[width=\columnwidth]{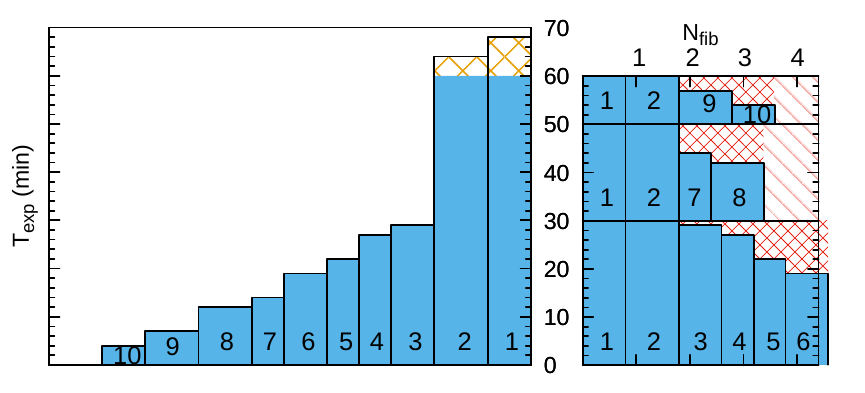}
        \caption{Schematic overview of the simplified fibre-to-target assignment for the 4MOST surveys. The left panels show LR targets in a small sky area ordered based on the requested exposure times. On average, this sky area includes 4.3 LR fibres (maximum $N_\mathrm{fib}$ value in the right-hand side panels). The column widths give the target completion fractions $f_\mathrm{compl}$. The column heights are the requested exposure times. The right panels show the distribution of these targets divided between three exposures with 30, 20, and 10~min, respectively. The division of targets between three exposures is done according to the algorithm described in Section~\ref{sec:Utargets}. The upper panels show nearly perfect fibre-to-target assignment. The lower panels show a fibre-to-target assignment for targets, where the target distribution does not allow for ideal allocation and some fibres are left empty. Different energy function components are shown with different colours and patterns.}
        \label{fig:fibtotar}
\end{figure}

The simplified fibre-to-target assignment is illustrated in Fig.~\ref{fig:fibtotar}. The upper panels show an almost perfect fibre allocation. The lower panels show target allocation in the case of a target distribution that does not allow an optimal fibre allocation. Regardless of the tile exposure times, some of the fibres are always empty, while long-exposure targets are not fully observed. This situation can only be improved by changing the target distribution in the sky. The proposed tiling algorithm tries to minimize the time that is missing and the time that is not used.

After the simplified fibre-to-target assignment, we have for each target $t\in\mathbf{t}_s^\mathrm{LR}$ the completion fraction $f_\mathrm{obs}(t)$ and for each tile $y\in\mathbf{y}_s$ we have the number of allocated fibres $N_\mathrm{fib}^\mathrm{alloc}(y)$. Additionally, if target $t$ was over-exposed ($f_\mathrm{obs}(t)>1.0$) we have the over-exposed fraction $f_\mathrm{over-exp}(t)$.

\paragraph*{Calculation of  $T^s_\mathrm{over-exp}$.} Over-exposed time in a region $s$ is estimated using the over-exposed fraction of each target
\begin{equation}
    T^s_\mathrm{over-exp,LR} = \frac{ 1 }{N^s_\mathrm{LR,fib}} \sum\limits_{t\in \mathbf{t}_s^\mathrm{LR}} f_\mathrm{over-exp}(t) \cdot T_\mathrm{exp}^\mathrm{D}(t) \cdot f_\mathrm{compl}(t) \,,
    \label{eq:texp_overexp}
\end{equation}

\paragraph*{Calculation of  $T^s_\mathrm{not-used}$.} The total time for empty fibres is estimated as
\begin{equation}
    T^s_\mathrm{not-used} = \frac{ 1 }{N^s_\mathrm{LR,fib}} \sum\limits_{y\in\mathbf{y}_s} \left\{\max\left[0,N^s_\mathrm{LR,fib} - N_\mathrm{fib}^\mathrm{alloc}(y) \right]  \right\}\cdot T_\mathrm{exp}^y,
\end{equation}
where $T_\mathrm{exp}^y$ is the exposure time of tile $y$.

\subsubsection{$U_\mathrm{overhead}(\mathbf{y}\,|\,\theta)$}

The energy function component for overheads, $U_\mathrm{overhead}(\mathbf{y}\,|\,\theta)$, is introduced to reduce the total amount of overhead time that is associated with each observation. For each science exposure, there is additional $T^\mathrm{tile}_{\mathrm{overhead}}=4.4$~min overhead (including calibration). Hence, for short exposures the fractional overhead is larger than for long exposures. At the same time, there are many targets that require short exposures, hence, short exposures are the optimal in some sky regions. In addition to the overhead associated with each exposure, there is additional overhead $T^\mathrm{OB}_{\mathrm{overhead}}=3.5$~min associated with each OB. This mainly covers the time for the telescope to move from one sky region to the other and thereafter acquiring the necessary guide stars. To reduce the summed  $T^\mathrm{OB}_{\mathrm{overhead}}$, 4MOST will combine several exposures into one OB.

The energy function that minimises the overhead time is defined as:
\begin{equation}
    U_\mathrm{overhead}(\mathbf{y}\,|\,\theta) = c_\mathrm{overhead}\left[ N_\mathrm{tile} T^\mathrm{tile}_{\mathrm{overhead}} + N_\mathrm{OB} T^\mathrm{OB}_{\mathrm{overhead}}\right],
    \label{eq:overhead}
\end{equation}
where $N_\mathrm{tile}$ is the total number of tiles (exposures) and $N_\mathrm{OB}$ is the number of individual OBs in the tiling solution. The parameter $c_\mathrm{overhead}$ can be used to fine-tune the importance of overhead energy component in the optimisation process.

\subsubsection{$U_\mathrm{tiles}(\mathbf{y}\,|\,\theta)$}
\label{sec:Utiles}

One 4MOST field of view is a hexagon. If we have to cover sky only once, then the optimal tiling is a beehive pattern. If some sky regions should be observed many times, then the optimal pattern should follow the target distribution in the sky. In intermediate cases, where the sky should be covered only twice, the optimal tiling is a beehive pattern that covers the sky twice. Since there is a small overlap between neighbouring tiles (because of the curved spherical surface of the sky), the two beehive patterns should be shifted with respect to each other, which minimises the number of overlaps in any sky location. To encourage this kind of pattern, the energy function $U_\mathrm{tiles}(\mathbf{y}\,|\,\theta)$ is defined as
\begin{eqnarray}
    U_\mathrm{tiles}(\mathbf{y}\,|\,\theta) = c_\mathrm{tiles} \cdot \qquad\qquad\qquad\qquad\qquad\qquad \nonumber \\
    \sum\limits_{i=1}^{N_\mathrm{OB}}\left\{
    R_{\lim} - \min\left[R_{\lim}, d(y^\mathrm{OB}_i,y_{k\neq i}^\mathrm{OB}:k=1\dots N_\mathrm{OB} ) \right] \right\},
    \label{eq:Utiles}
\end{eqnarray}
where $d(y^\mathrm{OB}_i,y_k^\mathrm{OB})$ is the angular distance between two OB centres and $R_{\lim}$ is the limiting radius. If the distance between OBs is larger than $R_{\lim}$ then there is no penalty in the energy function. If the distance between two OBs is smaller than $R_{\lim}$, we add a small penalty to the total energy $U(\mathbf{y}\,|\,\theta)$. Optimal $R_{\lim}$ should be close to the radius of one field of view, which also minimises the gaps between tiles. The interaction between tiles defined with Eq.~\eqref{eq:Utiles} is known as a nearest neighbour interaction in the point processes applications \citep[see e.g.][]{vanLieshout:00}.

If required by the surveys, a similar scheme can be used to force gaps between OBs in some specific sky areas. In this case the $U_\mathrm{tiles}(\mathbf{y}\,|\,\theta)$ should be defined individually for each sky region. In the current paper, for simplicity, we only test the energy function given with Eq.~\eqref{eq:Utiles}.

\subsubsection{$U_\mathrm{BGD}(\mathbf{y}\,|\,\theta)$}

The energy function component $U_\mathrm{targets}(\mathbf{y}\,|\,\theta)$ depends on whether a given sky region is observed during bright, grey or dark sky brightness conditions. Since observations during dark time are generally preferred, the previously introduced energy function components highly prefer observations during dark time. However, the fraction of total bright, grey and dark time is fixed. To take that into account, we introduce an energy function component $U_\mathrm{BGD}(\mathbf{y}\,|\,\theta)$ that somewhat balances the total time between B/G/D.

This energy function component is defined as
\begin{equation}
    U_\mathrm{BGD}(\mathbf{y}\,|\,\theta) =
    c_\mathrm{B}N^\mathrm{B}_\mathrm{tile} + c_\mathrm{G}N^\mathrm{G}_\mathrm{tile} + c_\mathrm{D}N^\mathrm{D}_\mathrm{tile} ,
    \label{eq:Ubgd}
\end{equation}
where $N^\mathrm{B/G/D}_\mathrm{tile}$ is the number of tiles with B/G/D flag and $c_\mathrm{B/G/D}$ are constants. In practice, $c_\mathrm{D}>c_\mathrm{G}>c_\mathrm{B}$, which slightly encourages bright and grey time tiles over dark time tiles. The parameters $c_\mathrm{B/G/D}$ should be chosen so that the fraction of the total bright, grey and dark time is as expected.

\subsection{Simulation method}
\label{sec:simulation}

To simulate marked point processes, several techniques can be used: spatial birth-and-death processes, Metropolis-Hastings (MH) algorithms, reversible jump dynamics or exact simulation techniques \citep{Geyer:94, Green:95, Geyer:99, Kendall:00, vanLieshout:00, vanLieshout:06, vanLieshout:19}.

In the current paper, we need to sample from the law $p(\mathbf{y}\,|\,\theta)$. This is done by using an iterative Monte Carlo algorithm. In our case the model parameters $\theta$ are fixed and conditional on $\theta$, and the object pattern is sampled from $p(\mathbf{y}\,|\,\theta)$ using an MH algorithm \citep{Geyer:94, Geyer:99}. The MH algorithm in this paper consists of three types of moves.

(\emph{i}) \textbf{Birth:} with a probability $p_\mathrm{b}$ a new object $\zeta$, sampled from the birth rate $b(\mathbf{y},\zeta)$, is proposed to be added to the present configuration $\mathbf{y}$. The new configuration $\mathbf{y}\prime = \mathbf{y}\cup\zeta$ is accepted with the probability
\begin{equation}
        \min\left\{ 1,\frac{p_\mathrm{d}}{p_\mathrm{b}}
        \frac{d(\mathbf{y}\cup\zeta,\zeta)}{b(\mathbf{y},\zeta)}
        \frac{p(\mathbf{y}\cup\zeta)}{p(\mathbf{y})} \right\}.
        \label{eq:pbirth}
\end{equation}

(\emph{ii}) \textbf{Death:} with a probability $p_\mathrm{d}$ an object $\zeta$ from the current configuration $\mathbf{y}$ is proposed to be eliminated according to the death proposal $d(\mathbf{y},\zeta)$. The probability of accepting the new configuration $\mathbf{y}\backslash \zeta$ (the set of objects $\mathbf{y}$ omitting the object $\zeta$) is computed by inverting the ratio~(\ref{eq:pbirth}).

(\emph{iii}) \textbf{Change:} with a probability $p_\mathrm{c}$ we randomly choose an object $\zeta_\mathrm{old}$ in the configuration $\mathbf{y}$ and propose to slightly change its parameters using uniform proposals. The new object obtained is $\zeta_\mathrm{new}$. The new configuration $\mathbf{y}\prime = \mathbf{y}\setminus\zeta_\mathrm{old}\cup\zeta_\mathrm{new}$ is accepted with the probability $\min\{ 1, p(\mathbf{y}\prime)/p(\mathbf{y}) \}$.

\begin{figure}
\centering
\includegraphics[width=\columnwidth]{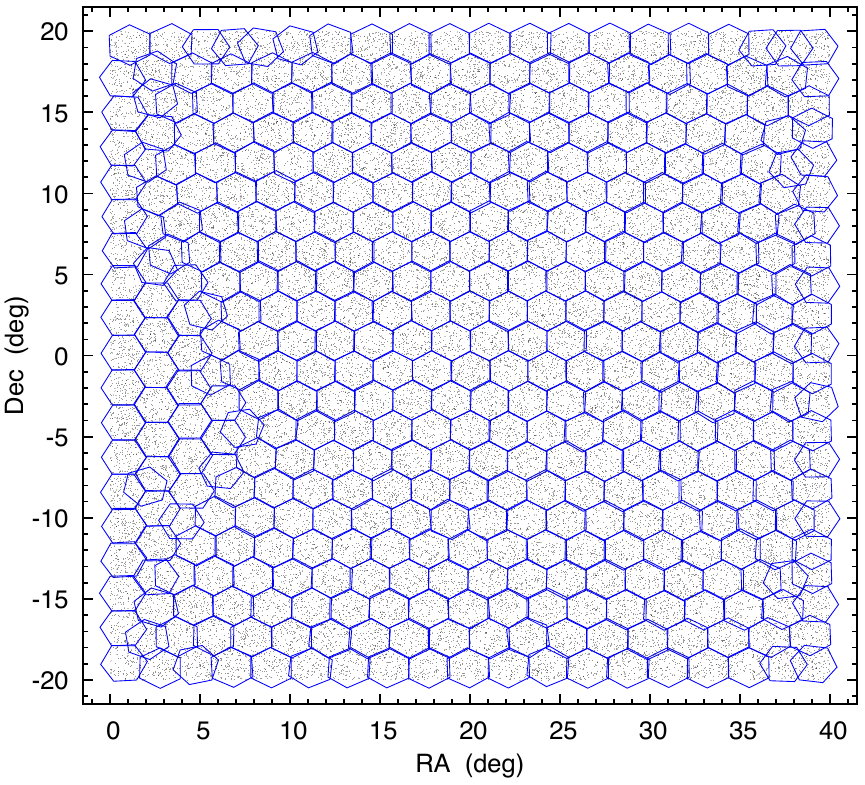}
        \caption{Optimal tiling pattern in the case of one visit. Each tile is shown as a blue hexagon. Uniformly distributed targets (grey dots) are restricted in the right ascension range $0\dots40$~deg and declination range $-20\dots+20$~deg. For clarity, only 10 per cent of the targets are shown. Stitches in the tiling pattern are due to the hexagons along the neighbouring edges being rotated by 90 degrees.}
        \label{fig:tiles_one_visit}
\end{figure}

\begin{figure*}
\centering
\includegraphics[width=0.99\textwidth]{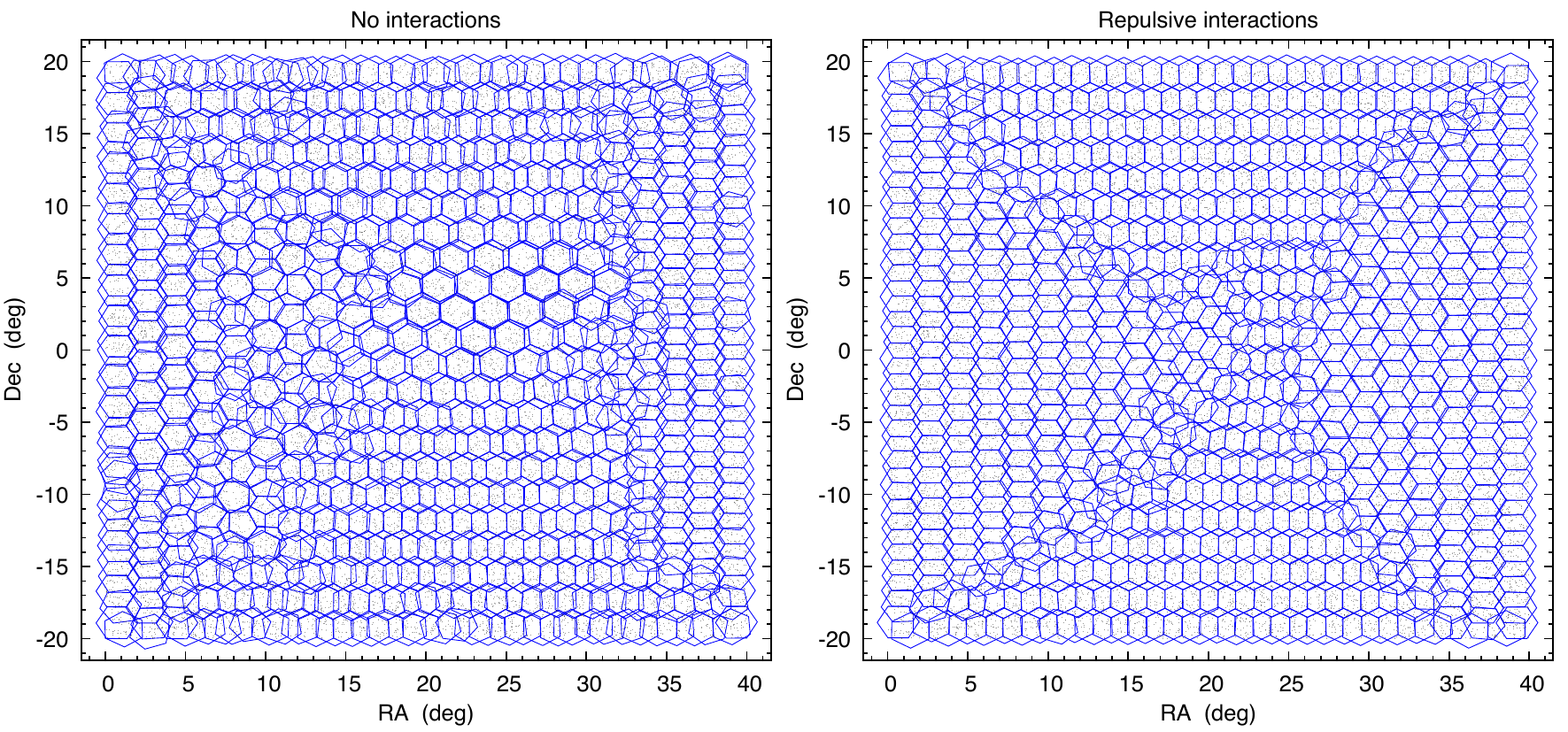}
        \caption{Optimal tiling pattern in the case of two visits. Each tile is represented as a blue hexagon. Targets are restricted in the same area as in Fig.~\ref{fig:tiles_one_visit}. On the left-hand panel there are no interactions between tiles, $U_\mathrm{tiles}=0.0$. On the right-hand panel we use the repulsive interactions (see Section~\ref{sec:Utiles}) where tile centres are maximally pushed apart from each other. Clearly, the resulting tiling pattern depends on the type of interactions among tiles.}
        \label{fig:tiles_two_visit}
\end{figure*}

\begin{figure}
\centering
\includegraphics[width=\columnwidth]{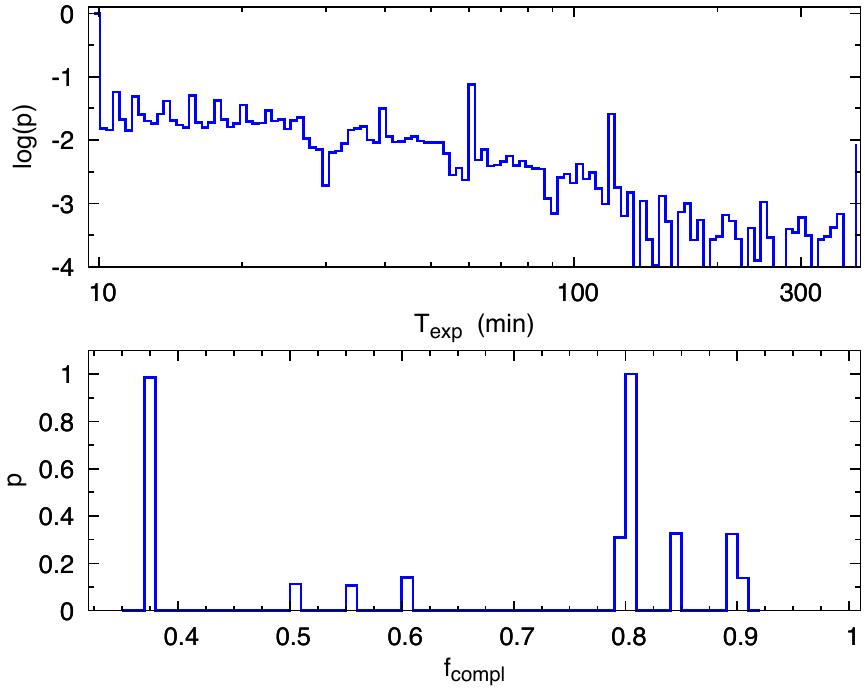}
        \caption{The distribution of exposure times (upper panel) and $f_\mathrm{compl}$ values (lower panel) for the targets shown in Fig.~\ref{fig:sky_pixels}. The distribution is shown in arbitrary units. The minimum exposure time for all targets was set to 10 minutes. Most of the long-exposure targets are located in the WAVES region (see Fig.~\ref{fig:sky_pixels}). In the input catalogue, each sub-survey has a fixed $f_\mathrm{compl}$ value. The variety of $f_\mathrm{compl}$ values shows that the completeness requirements in various surveys are very different.}
        \label{fig:test_targets}
\end{figure}

For the death rate, we adopt the uniform choice $d(\mathbf{y},\zeta) = 1/n(\mathbf{y})$, where $n(\mathbf{y})$ is the number of objects in the configuration. For the birth proposal above, we have a mixture proposal with two types of sub-moves:
\begin{itemize}
    \item \textbf{Random:} with a probability $p^\mathrm{rnd}_\mathrm{b}$ a new random tile (a new OB with just one tile) is added to the configuration. The tile centre is chosen uniformly in the sky (in the observed window $W$). For the new tile, we assign a random position angle and a random exposure time between $T_{\min}$ and $T_{\max}$. For the new tile we also attach a B/G/D flag, where the prior for the B/G/D flag is the fraction of time available in bright, grey or dark sky conditions.
    \item \textbf{Tile in an OB:} with a probability $p^\mathrm{OB}_\mathrm{b} = 1.0-p^\mathrm{rnd}_\mathrm{b}$ we choose an existing tile $\zeta\prime$ and add a new tile $\zeta$ to this OB. The tile coordinates, position angle and B/G/D flag become the same as for the existing OB. The exposure time for the new tile is chosen randomly between $T_{\min}$ and $T_{\max}$.
\end{itemize}
The birth rate for the combined birth move is
\begin{equation}
    b(\mathbf{y},\zeta) = \frac{p^\mathrm{rnd}_\mathrm{b}\mathbbm{1}\!\left\{\zeta\in W\right\}}{\nu(W)} +
    p^\mathrm{OB}_\mathrm{b} \tilde{b}(\mathbf{y},\zeta) ,
    \label{eq:birth_rate}
\end{equation}
\begin{equation}
    \tilde{b}(\mathbf{y},\zeta) = \frac{1}{n(\mathbf{y})}
    \sum\limits_{\zeta\prime\in \mathbf{y}}
    \frac{\mathbbm{1}\!\left\{\zeta\in b(\zeta\prime,r)\right\}}{\nu\left[b(\zeta\prime,r)\cap W\right]},
\end{equation}
where $\nu(W)$ is the Lebesgue measure (sky area) of the observed window $W$, $b(\zeta\prime,r)$ is a ball centred in $\zeta\prime$ with radius $r$ in the sky, and $\mathbbm{1}\!\left\{\cdot\right\}$ is the indicator function. For simplicity, we set the area of the ball $\nu\left[b(\zeta\prime,r)\right]=1$ and $\nu(W) = N_\mathrm{expected}$, where $N_\mathrm{expected}$ is the expected number of tiles in the converged solution. The actual number of tiles may differ from the expected number and is mainly determined by the energy function $U_\mathrm{targets}$.

For the change move above, we adopt the following sub-moves:
\begin{itemize}
    \item \textbf{Position in the sky:} with a probability $p^\mathrm{pos}_\mathrm{c}$ we slightly shift an OB centre (where the selected tile is in) and position angle with respect to the OBs original values.
    \item \textbf{Exposure time:} with a probability $p^\mathrm{exp}_\mathrm{c}$ we slightly change tile exposure time with respect to the original exposure time.
    \item \textbf{Change B/G/D flag:} with a probability $p^\mathrm{BGD}_\mathrm{c}$ we propose to change the B/G/D flag for a selected OB.
    \item \textbf{Combine close tiles into the same OB:} with a probability $p^\mathrm{OB}_\mathrm{c}$ we select randomly a tile and if there is another tile close to the selected tile, we join both tiles into one OB.
\end{itemize}
The previously introduced birth, death and change moves define a Markov chain transition kernel which is $\phi-$irreducible, Harris recurrent and geometric ergodic \citep{vanLieshout:00,MollWaag:04,Stoica:05}.

In order to maximise $p(\mathbf{y},\theta)$, the previously described sampling mechanism is integrated into a simulated annealing (SA) algorithm. The SA is an iterative algorithm that samples from $p(\mathbf{y},\theta)^{1/T}$, while $T$ goes slowly to zero. The following ingredients are needed in order to ensure convergence of the SA algorithm: high value of the initial temperature, a convergent sampling algorithm for the probability density and an appropriate cooling schedule \citep{vanLieshout:00,Stoica:05}. We adopt the polynomial cooling schedule, where the temperature is lowered as
\begin{equation}
    T_{k+1} = \alpha T_k,
    \label{eq:cooling}
\end{equation}
where $k$ is a time step in a simulation and $0<\alpha<1.0$ defines the speed of temperature decrease. The initial temperature for the simulation is set to $T_0$. The temperature is lowered after every $N_\mathrm{moves}$, which allows the system to reach a near-equilibrium state. In practice, the $N_\mathrm{moves}$ should be several times greater than the number of tiles in the configuration. Altogether we change the temperature $N_\mathrm{cycles}$ time. Hence, the total number of moves in our algorithm is $N_\mathrm{moves}\cdot N_\mathrm{cycles}$.

The MH algorithm described above does not require any tiling initialisation. The MCMC algorithm starts with zero tiles and additional tiles are added to the tiling configuration during birth moves. The final number of tiles in the configuration is mostly determined by the $U_\mathrm{targets}$ energy function and is influenced by the expected number of tiles, $N_\mathrm{expected}$.

\section{Application of the tiling algorithm}
\label{sec:application}

\subsection{Example using Poisson-distributed targets}

\begin{figure*}
\centering
\includegraphics[width=\textwidth]{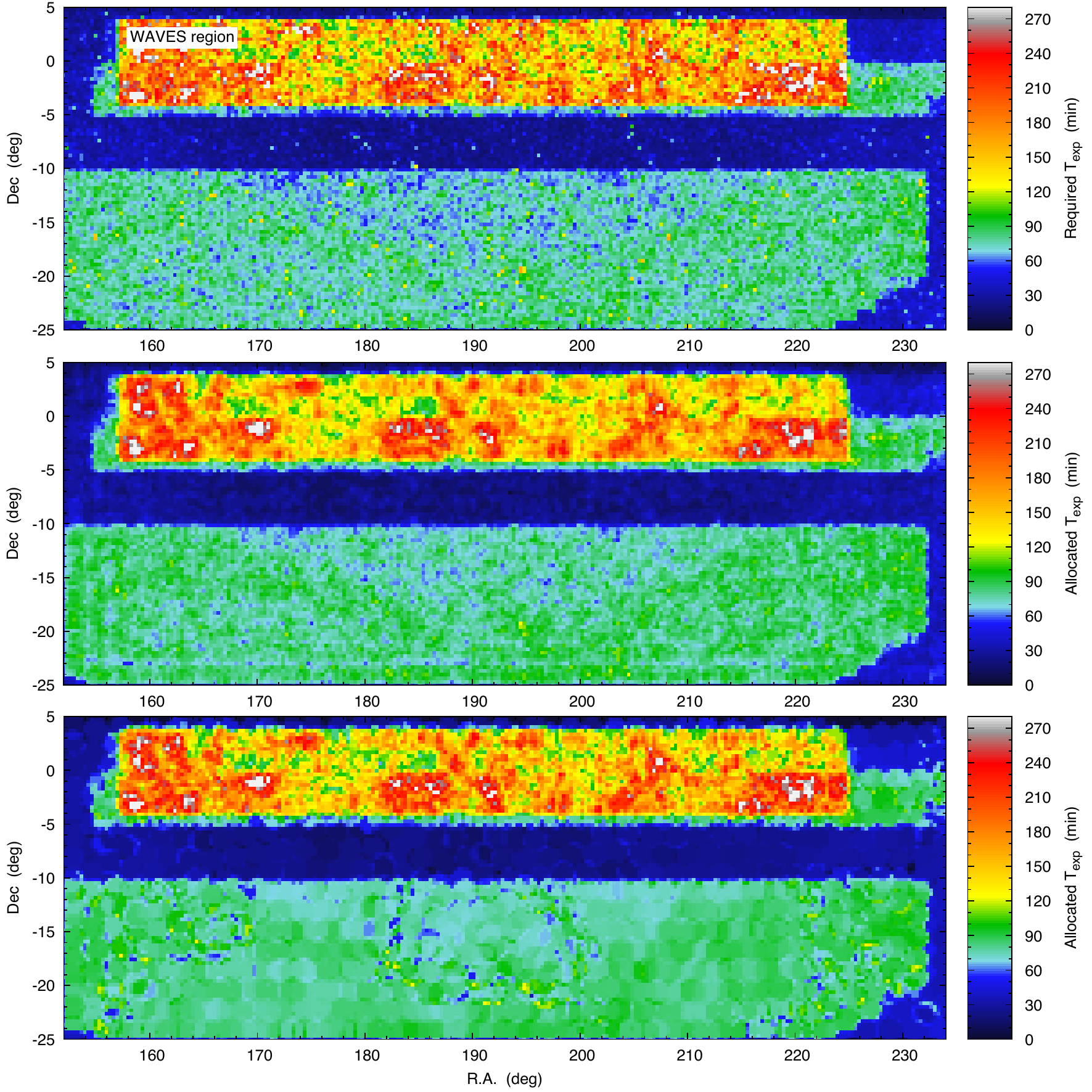}
        \caption{The upper panel shows the required exposure time for different sky regions. Targets for the test region were selected from the 4MOST mock catalogues. The required exposure time was calculated using Eq.~\protect\eqref{eq:texp_req}. The upper region with a high number density of targets is the WAVES survey region \citep{2019Msngr.175...46D}. Middle and lower panels show the allocated exposure times (sum of tiles exposures times) for the same test region. In the middle panel, each tile is an individual OB, while in the lower panel, tiles are collected into OBs, to reduce the total overhead time. In both cases, the allocated exposure time traces the required exposure time shown on the upper panel very well. Fig.~\protect\ref{fig:sky_tiling} gives the actual tiling pattern in the sky for these two cases. Fig.~\protect\ref{fig:sky_pixels_pot} shows how well the allocated exposure time matches with the required exposure time, while taking the fibre-to-target assignment into account.}
        \label{fig:sky_pixels}
\end{figure*}

\begin{figure*}
\centering
\includegraphics[width=\textwidth]{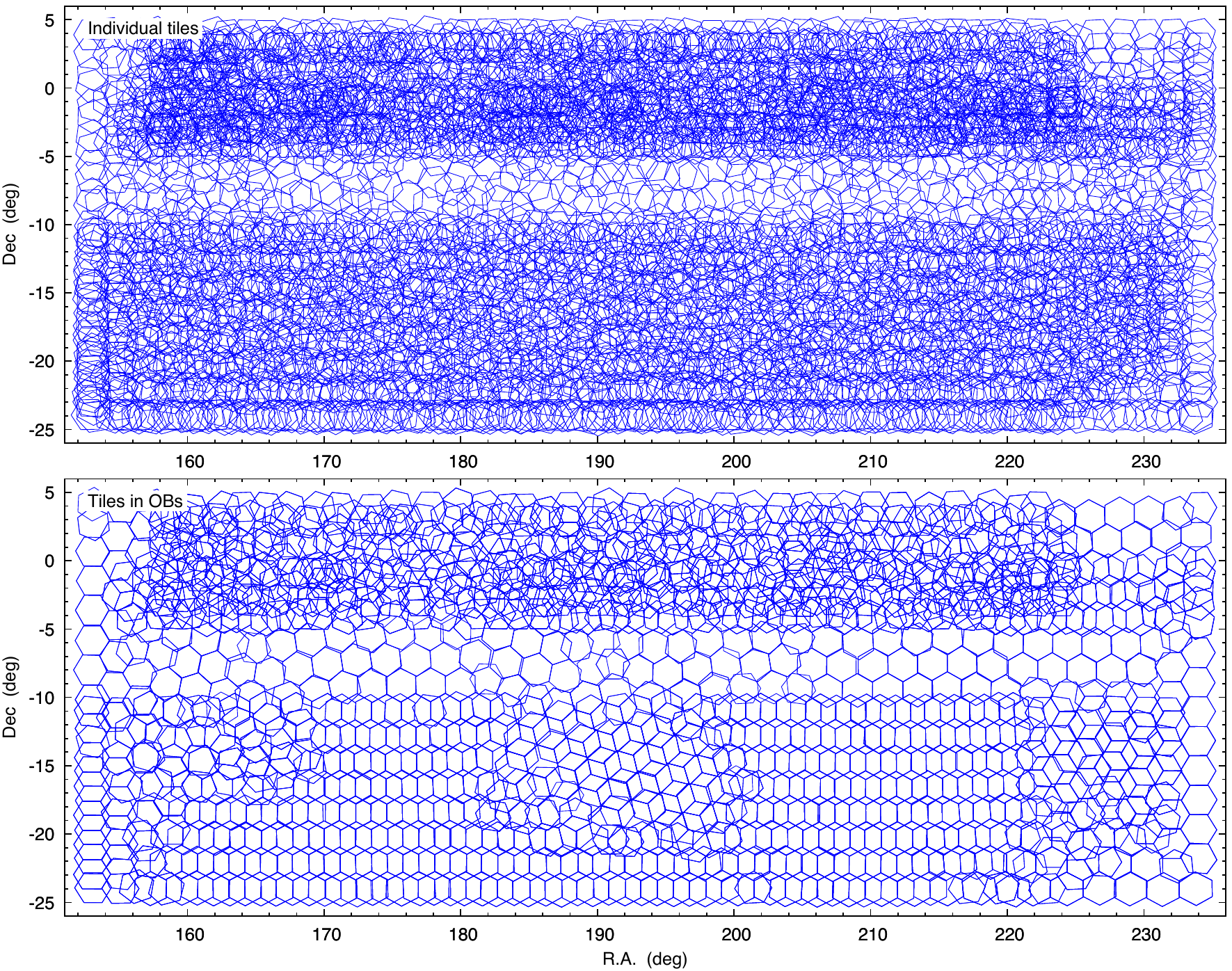}
        \caption{Tiling pattern in the sky that corresponds to the allocated exposure times shown in the middle and lower panel in Fig.~\ref{fig:sky_pixels}. Each OB is shown as a blue hexagon. In the upper panel, each tile is an individual OB. In the lower panel, several tiles are collected into single OBs. Clearly, the flexibility of the tiling pattern depends on the number of OBs.}
        \label{fig:sky_tiling}
\end{figure*}

\begin{figure*}
\centering
\includegraphics[width=\textwidth]{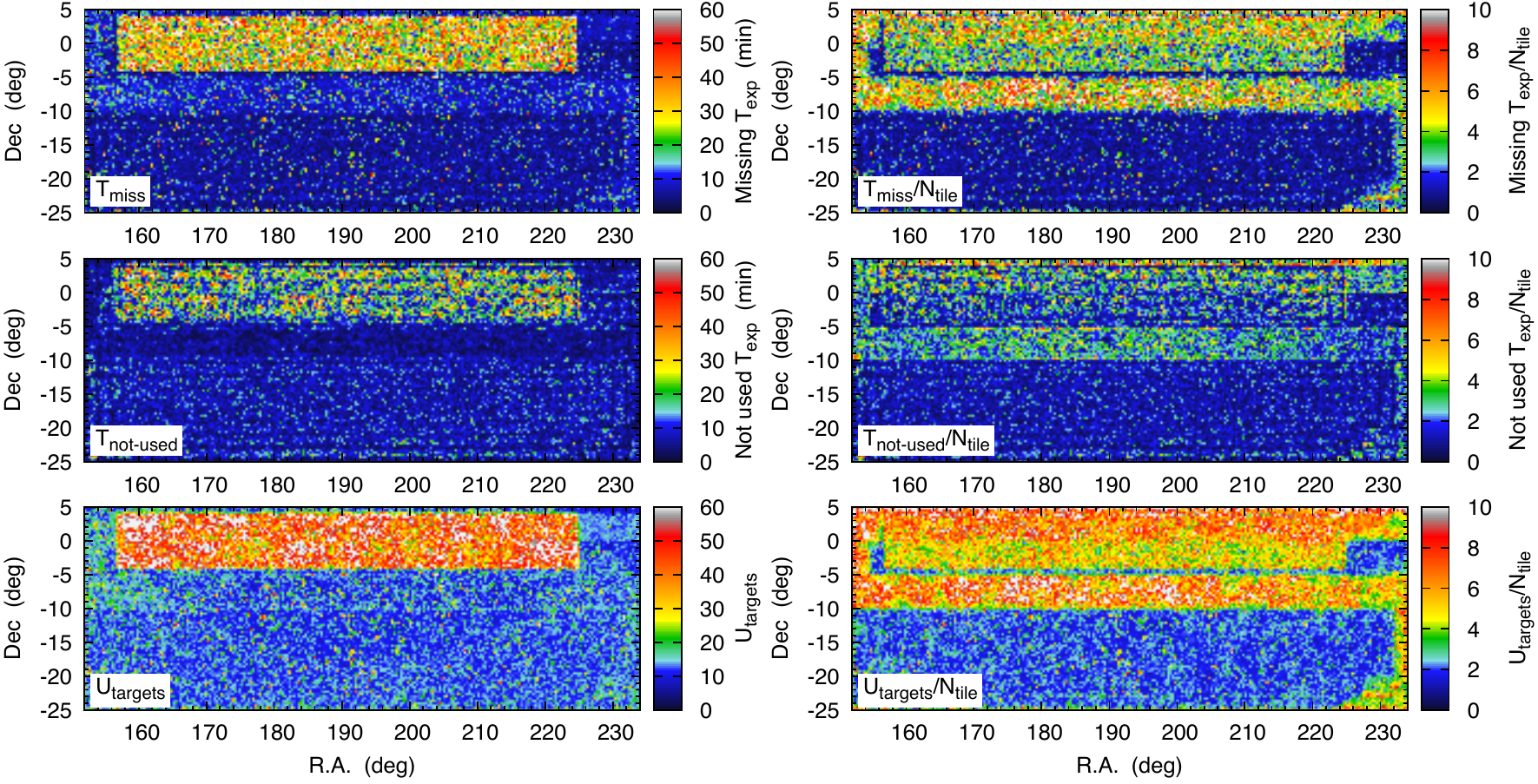}
        \caption{Diagnosis plots of the tiling algorithm. The left-hand column shows the energy function $U_\mathrm{targets}$ (lower panel) and its components $T_\mathrm{miss}$ (upper panel) and $T_\mathrm{not-used}$ (middle panel). In general, $T_\mathrm{miss}$ counts the time that is missing to observe the required set of targets and $T_\mathrm{not-used}$ counts the time that is wasted because of empty fibres. The $U_\mathrm{targets}$ is the combination of these (see Section~\ref{sec:Utargets}). The right-hand panels show the same energy function components divided by the number of tiles in a given sky region. This Figure shows the energy function components for the tiling presented in the middle panel of Fig.~\ref{fig:sky_pixels}.}
        \label{fig:sky_pixels_pot}
\end{figure*}

\begin{figure*}
\centering
\includegraphics[width=\textwidth]{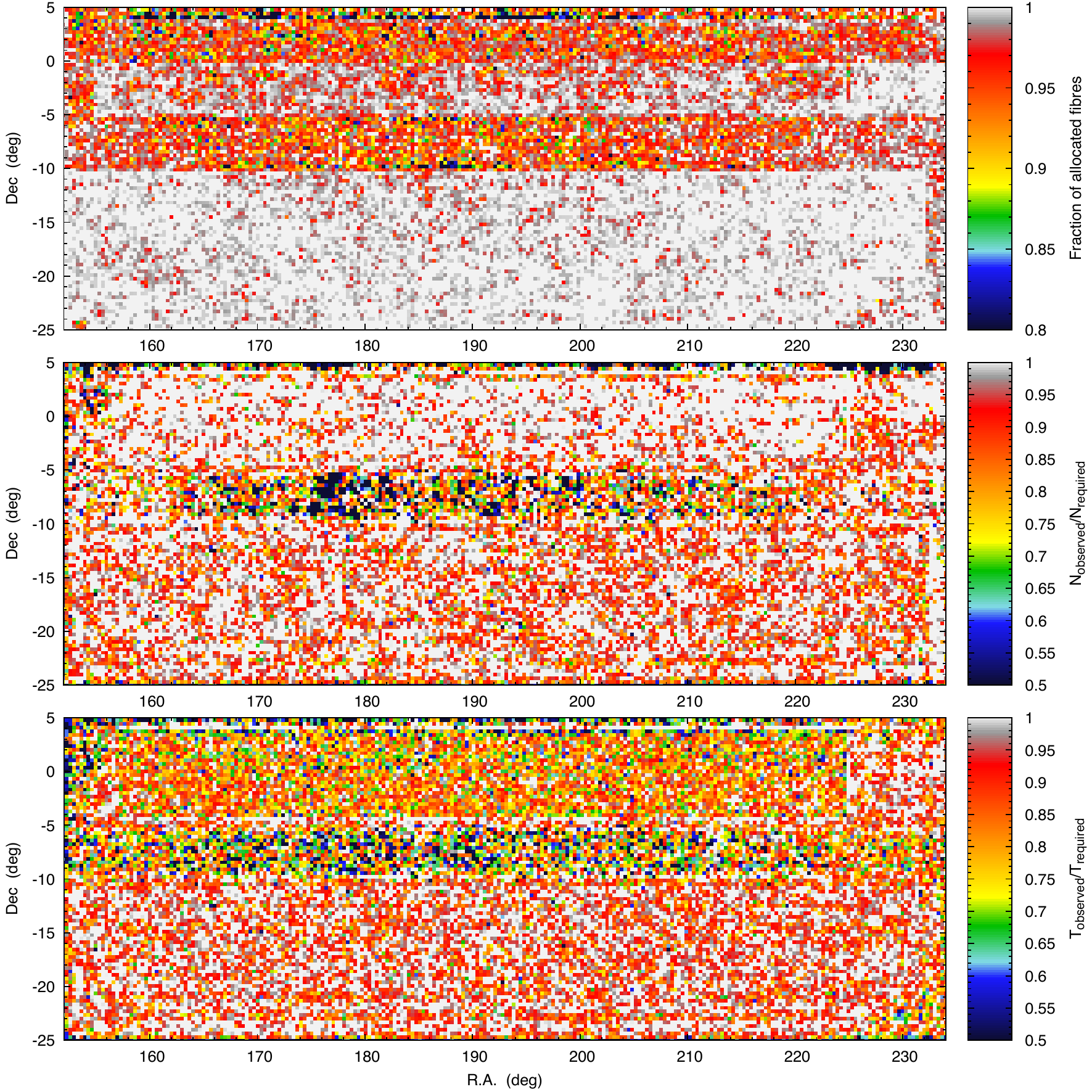}
        \caption{Output of probabilistic fibre-to-target assignment. Using the mock targets shown in the top panel of Fig.~\ref{fig:sky_pixels} and tiling map shown in the upper panel of Fig.~\ref{fig:sky_tiling}, we ran the survey simulation using the probabilistic fibre-to-target assignment presented in \citet{Tempel:2019}. See Section~\ref{sec:prob_sim} for more details. The upper panel shows the fraction of allocated fibres. The middle panel shows the number of successfully observed objects divided by the number of required objects. The lower panel shows the fraction of observed exposure time out of required exposure time.}
        \label{fig:sky_pixels_sim}
\end{figure*}

As a first test, we generated Poisson-distributed targets in the sky and ran the tiling algorithm on these points. The targets were restricted in the right ascension range $0\ldots40$~deg and declination range $-20\ldots +20$~deg. The number density of targets was slightly lower than the number density of fibres. Hence, the expected optimal tiling pattern covers the sky only once.

Fig.~\ref{fig:tiles_one_visit} shows the optimal tiling pattern generated using the algorithm described in Section~\ref{sec:point_process}. An ideal theoretical tiling pattern would be a perfect honey-comb pattern. However, the celestial sphere is curved, the target region in the sky is restricted and the tiles should not be located outside of the target region, so the perfect honey-comb pattern cannot be achieved exactly. The stitches visible in Fig.~\ref{fig:tiles_one_visit} are due to the orientation of tiles (hexagons) at the neighbouring edges of the sky area being rotated by 90 degrees. The orientation of tiles at the field edges are determined by the sharp edge of the field and fixed hexagon orientations are the only solution we found to produce an optimal tiling that minimises the tile area outside of the target region. For a large field of view, these stitches are not present and the algorithm generates a nearly perfect honey-comb pattern in the case of one visit to each tile.

As a second test with Poisson-distributed targets, we doubled the number of targets. Consequently, the expected perfect tiling covers the sky twice. We use this test to show the effect of $U_\mathrm{tiles}$ (see Section~\ref{sec:Utiles}) on the final tiling pattern. In Fig.~\ref{fig:tiles_two_visit} we show the tiling pattern generated in two different cases. In the left panel, we show the tiling pattern where $U_\mathrm{tiles}=0.0$, which means that there are no interactions between tiles. In the right-hand panel of Fig.~\ref{fig:tiles_two_visit} we show the tiling pattern where the repulsive interaction of tiles is added, as described in Section~\ref{sec:Utiles}. The parameter $c_\mathrm{tiles}=5.0$, which is relatively large to forces the tile centres maximally apart from each other.

Fig.~\ref{fig:tiles_two_visit} clearly shows that the final tiling pattern depends on the choice of $U_\mathrm{tiles}$. In the left-hand panel, in several locations two tiles are put almost perfectly on top of each other. In the right-hand panel, the distance between tile centres is maximised and the resulting pattern appears more regular. In both cases, the number of tiles is practically the same and both patterns cover the sky twice with minimal overlaps and holes between tiles. Hence, $U_\mathrm{tiles}$ has a negligible effect on the survey efficiency. In the proposed algorithm, $U_\mathrm{tiles}$ can be used to influence the tiling pattern so that it maximises the survey science goals. Depending on the survey, these goals can be rather different.

The tiling solutions presented in Figs.~\ref{fig:tiles_one_visit} and \ref{fig:tiles_two_visit} each show just one realisation of a solution to the optimal tiling problem. Since the MCMC algorithm involves randomness, if we run the tiling algorithm a second time with exactly the same parameters and different random seeds, or using parallel computation as typically implemented (which is what we currently have coded), the outcome will be slightly different. For example, in Fig.~\ref{fig:tiles_one_visit}, the stitches will appear in different locations. Due to the complexity of the optimal tiling problem, it is hard to define the optimal tiling pattern. In practice, there are many optimal tiling patterns and the proposed algorithm only provides one numerical realisation of a solution to the problem. Due to the high-dimensionality of the problem, there are many local minima that are all approximately equal in practice and the MCMC algorithm provides one local minimum as a final solution. The optimal tiling pattern is defined via the energy function in Eq.~\eqref{eq:energy_function} and depends on its form and parameters. For different scientific applications the optimal tiling pattern might be different and the proposed algorithm allows to take this into account.

\subsection{Example using mock catalogues}
\label{sec:mock}

In this section, we test the proposed tiling algorithm in the case of a varying number density of targets in the sky. Targets are taken from the 4MOST mock catalogues, covering the Galactic and extragalactic consortium surveys (see Section~\ref{sec:intro}). The distribution of exposure times and $f_\mathrm{compl}$ values for targets in our test region are shown in Fig.~\ref{fig:test_targets}. The upper panel in Fig.~\ref{fig:sky_pixels} shows the required exposure time in a test sky region. The required exposure time is estimated with Eq.~\eqref{eq:texp_req}. Fig.~\ref{fig:sky_pixels} shows the footprints of individual surveys in the sky. The upper part of this Figure shows the WAVES survey where the number density of objects, as well as the required exposure time, varies significantly even on small scales.

To generate the tiling for the selected test region, we restricted ourselves to only LR targets and all tiles had the same sky brightness condition. We generated the tiling pattern for two cases. In the first case we set $U_\mathrm{overhead}=0.0$, so that each tile was considered to be an individual OB with no penalty from the overhead during tile generation. In the second case, we minimised the overhead associated with each exposure and OB, we set $c_\mathrm{overhead}=0.5$, and tiles were collected into OBs wherever possible and efficient.

The middle and lower panels in Fig.~\ref{fig:sky_pixels} show the allocated exposure time for these two cases. In general, the allocated time in different sky regions is roughly the same for both cases. In the lower part of the test region, the required exposure time is lower and there is less flexibility there. Hence, a slight hint of the tiling pattern of individual hexagons can be seen in the allocated exposure time shown in the lower panel of Fig.~\ref{fig:sky_pixels}. This is because tiles are collected into OBs and this part of the region is mostly covered only twice. In general, we can see that the allocated exposure time matches the required exposure times very well. In many cases, even the small variations in the required exposure time maps are well traced by the allocated exposure times. The required and allocated exposure times are not directly matched in our algorithm. During the optimisation, we minimise the missing and wasted time (see Section~\ref{sec:Utargets}), which automatically results in an excellent match between the required and allocated exposure times.

Fig.~\ref{fig:sky_tiling} shows the actual tiling pattern for the two cases shown in the lower two panels of Fig.~\ref{fig:sky_pixels}. This clearly emphasises that if tiles are collected into OBs, then we lose some flexibility when placing tiles in the sky. This flexibility comes with the price of increased overhead time. For this example, the summed exposure time without the overhead time for the two cases is nearly identical, being around 860 hours. However, the overhead for the first case is 446 hours, while for the second case it is 339 hours. Hence, there is a balance between an efficient survey (minimised overhead time) and an optimal tiling pattern (flexibility of placing tiles). This balance depends on the survey science goals and required completeness for a survey. The best compromise should be determined during the survey optimisation.

Fig.~\ref{fig:sky_pixels_pot} shows the energy function $U_\mathrm{targets}$ and different components of $U_\mathrm{targets}$. Clearly, the WAVES region has the highest energy, as it is the least efficient part of the selected test region, while at the same time it has the largest $T_\mathrm{missing}$ and $T_\mathrm{not-used}$ times. This is because of the nature of the WAVES region. The target density varies significantly at scales smaller than one 4MOST field of view. At the same time, the exposure times of individual targets differ a lot. As a consequence, there are regions where the allocated exposure time is smaller than the required exposure time for single targets and in the same regions there are empty fibres. A similar case is illustrated in the lower panel of Fig.~\ref{fig:fibtotar}. The algorithm minimises the sum of the two components. Depending on the science goals, the relative importance of the two components can be altered. For example, WAVES requires high completeness for its science case, so $T_\mathrm{missing}$ should be more important than $T_\mathrm{not-used}$. In the proposed algorithm, these survey-specific requirements can be easily included, while generating the final optimal tiling for the 4MOST survey.

Regarding the distribution of missing and not-used time in the sky and the distribution of energy function $U_\mathrm{targets}$, these are more or less uniform outside and inside the WAVES region. The proposed tiling algorithm finds the tiling that matches the required exposure times and finds a solution, where missing and not-used time is evenly distributed in the sky. Hence, the proposed tiling algorithm does not seemingly prefer one sky region to the other. The difference between WAVES and other regions is due to the different target densities, completeness requirements and exposure time distributions.

While the left-hand panels in Fig.~\ref{fig:sky_pixels_pot} show the energy function per sky region, the right-hand panels show the same energy function per tile, the energy function components are divided by the number of tiles in a given sky region. While the summed energy is the lowest in the middle of the region, the energy per tile is highest there. This is because the number density of objects there is relatively low and this region is covered mainly with one layer of tiles. Since there is almost no flexibility for the tiling pattern in this region, the survey efficiency largely depends on the match between the target density and fibre density. The mismatch between these two is the reason why the energy per tile is highest there. To conclude, the energy function maps are useful for analysing the overall efficiency of the generated tiling pattern. However, the tiling pattern should be used together with the real fibre-to-target assignment and the actual survey efficiency can be only assessed using the full survey simulation. This is briefly analysed in the next section.

\begin{figure*}
\centering
\includegraphics[width=\textwidth]{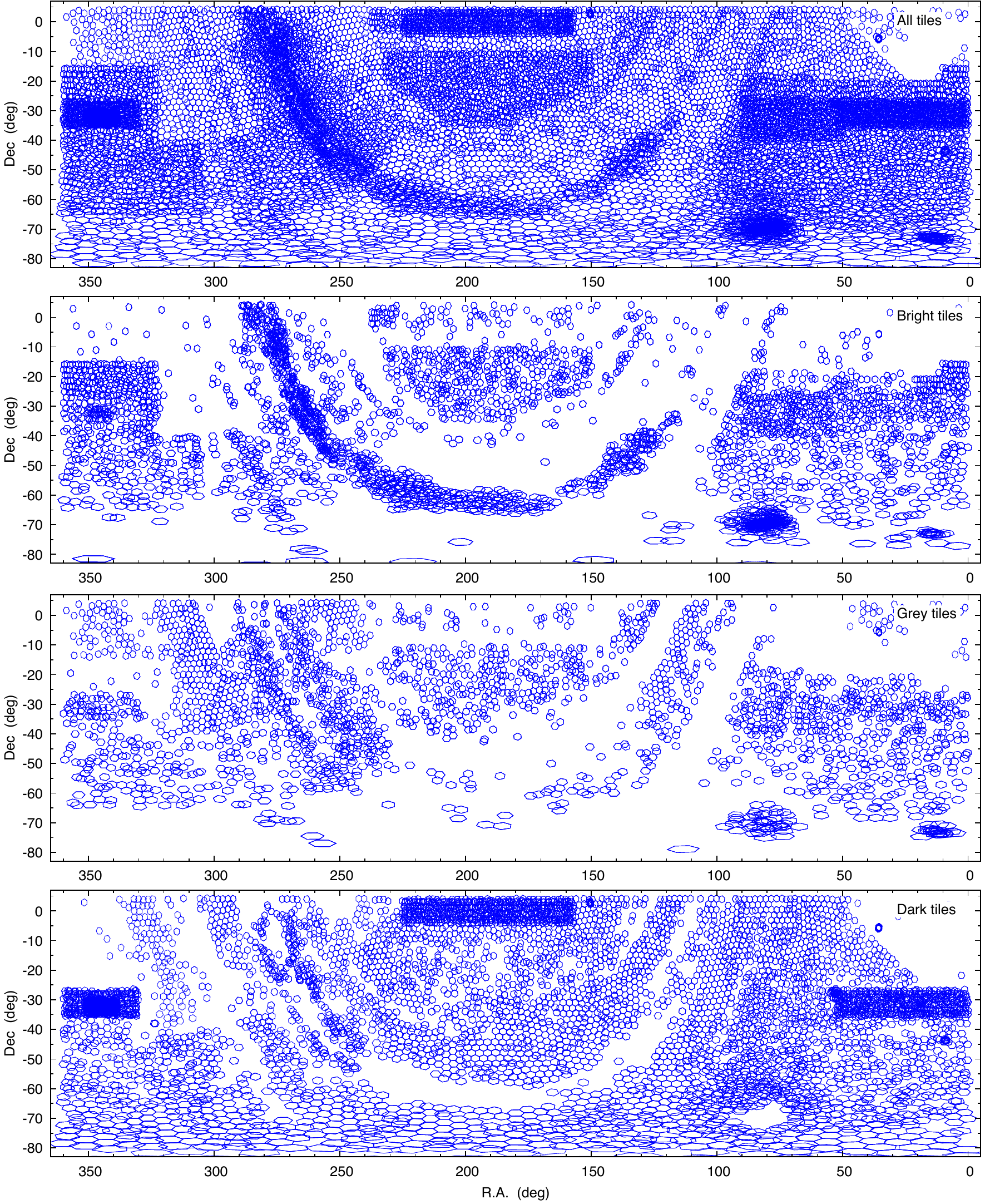}
        \caption{Tiling pattern for all 4MOST mock catalogues. The distribution of targets in the mock catalogues is shown in Fig.~\protect\ref{fig:texp_map}. The upper panel shows all tiles that are necessary to observe the required set of targets from the mock catalogues. The lower panels show the tiles for bright, grey and dark sky conditions. The division between different sky conditions works as expected. The Milky Way and the Magellanic Clouds are mostly observed during bright time, while the extragalactic sky is mostly observed during dark time.}
        \label{fig:tiling_fullsky}
\end{figure*}

\subsection{Tiling and probabilistic fibre-to-target assignment}
\label{sec:prob_sim}

The fibre-to-target assignment described in this paper is a simplified approach that provides only a statistical solution and cannot be used during real observations. In reality, the generated tiling pattern will be used together with a more sophisticated fibre-to-target assignment algorithm. In \citet{Tempel:2019} we proposed a probabilistic fibre-to-target assignment algorithm that takes into account survey completeness requirements and varying number densities of targets. In this section, we will adopt the tiling pattern generated in Section~\ref{sec:mock} and use this together with the probabilistic fibre-to-target assignment described in \citet{Tempel:2019}. For the probabilistic fibre-to-target assignment, we use the pattern shown in the upper panel of Fig.~\ref{fig:sky_tiling} as an input.

Fig.~\ref{fig:sky_pixels_sim} shows the efficiency of the probabilistic fibre-to-target assignment algorithm. The upper panel shows the fraction of allocated fibres. This is close to unity in the lower part of the Figure. In the upper part, the fraction of used fibres is on average greater than 95\,\%. The fraction of used fibres is lower than 90\,\% only in some small regions. The low efficiency is in regions where the completeness requirement is very high (WAVES region) or the number density of objects is low (middle region in the figure). In general, the adopted tiling pattern is not visible in the completeness map and the completeness differences are caused by the different number density of objects.

The middle and lower panels in Fig.~\ref{fig:sky_pixels_sim} show the fraction of successfully observed objects and of used exposure time compared with the required number of objects and exposure times. In most of the figure, both fractions are greater than 90\,\%. The greatest difference is in the WAVES region, where the number of observed objects is close to the number of required objects, while the fraction of used exposure time is lower. This is because in the WAVES region we observe more of the required short-exposure targets, while some long-exposure targets remain uncompleted (the total exposure time is shorter than the requested exposure time for a target). This situation can only be improved by making the tiling less efficient, while adding more tiles (the fraction of allocated fibres will decrease) or increasing the exposure time of tiles (the over-exposure of short-exposure targets will increase).

To summarise, the simplified fibre-to-target assignment used during the tiling pattern generation works well and is a good approximation of the probabilistic fibre-to-target algorithm presented in \citet{Tempel:2019}. Further improvements of the simplified fibre-to-target assignment algorithm should take into account survey-specific requirements. This will be done during the 4MOST survey optimisation phase.

\subsection{Tiling with bright, grey and dark time division}

\begin{table}
        \centering
        \caption{Parameter values in our tiling algorithm during test simulations. The last column gives the reference to the equation or section, where the parameter is used or discussed.}
        \label{tab:parameters}
        \begin{tabular}{lrll} 
                \hline
                Parameter & Value & Unit & Reference\\
                \hline
                $c_\mathrm{miss}$ & 1.0 & & Eq.~\eqref{eq:Us_targets}\\
                $c_\mathrm{wasted}$ & 0.5 & & Eq.~\eqref{eq:Us_targets}\\
                $s_\mathrm{max}$ & 0.1 & deg & Eq.~\eqref{eq:Us_targets}\\
                $c_\mathrm{LR}$ & 2/3 & & Eqs.~\eqref{eq:tmissing} and \eqref{eq:twasted}\\
                $c_\mathrm{HR}$ & 1/3 & & Eqs.~\eqref{eq:tmissing} and \eqref{eq:twasted}\\
                $\rho_\mathrm{fib}^\mathrm{LR}$ & 391 & & Eq.~\eqref{eq:texp_req}\\
                $\rho_\mathrm{fib}^\mathrm{HR}$ & 196 & & Eq.~\eqref{eq:texp_req}\\
                $c_\mathrm{sci\_fib}$ & 0.85 & & Eq.~\eqref{eq:texp_req}\\
                $c_\mathrm{overhead}$ & 0.5 & & Eq.~\eqref{eq:overhead}\\
                $T_\mathrm{overhead}^\mathrm{tile}$ & 4.4 & min & Eq.~\eqref{eq:overhead}\\
                $T_\mathrm{overhead}^\mathrm{OB}$ & 3.5 & min & Eq.~\eqref{eq:overhead}\\
                $c_\mathrm{tiles}$ & 2.0 & & Eq.~\eqref{eq:Utiles}\\
                $R_\mathrm{lim}$ & 0.8 & deg & Eq.~\eqref{eq:Utiles}\\
                $c_\mathrm{B}$ & 5.0 & & Eq.~\eqref{eq:Ubgd}\\
                $c_\mathrm{G}$ & 3.5 & & Eq.~\eqref{eq:Ubgd}\\
                $c_\mathrm{D}$ & 2.0 & & Eq.~\eqref{eq:Ubgd}\\
                $p_\mathrm{b}$ & 0.2 & & Sect.~\ref{sec:simulation} and Eq.~\eqref{eq:pbirth}\\
                $p_\mathrm{d}$ & 0.2 & & Sect.~\ref{sec:simulation} and Eq.~\eqref{eq:pbirth}\\
                $T_\mathrm{min}$ & 5 & min & Sect.~\ref{sec:simulation}\\
                $T_\mathrm{max}$ & 30 & min & Sect.~\ref{sec:simulation}\\
                $p_\mathrm{c}$ & 0.6 & & Sect.~\ref{sec:simulation}\\
                $p_\mathrm{b}^\mathrm{rnd}$ & 0.4 & & Sect.~\ref{sec:simulation} and Eq.~\eqref{eq:birth_rate}\\
                $p_\mathrm{b}^\mathrm{OB}$ & 0.6 & & Sect.~\ref{sec:simulation} and Eq.~\eqref{eq:birth_rate}\\
                $N_\mathrm{expected}$ & 30\,000 & & Sect.~\ref{sec:simulation} and Eq.~\eqref{eq:birth_rate}\\
                $p_\mathrm{c}^\mathrm{pos}$ & 0.3 & & Sect.~\ref{sec:simulation}\\
                $p_\mathrm{c}^\mathrm{exp}$ & 0.3 & & Sect.~\ref{sec:simulation}\\
                $p_\mathrm{c}^\mathrm{BGD}$ & 0.3 & & Sect.~\ref{sec:simulation}\\
                $p_\mathrm{c}^\mathrm{OB}$ & 0.1 & & Sect.~\ref{sec:simulation}\\
                $T_\mathrm{0}$ & 1.0 & & Sect.~\ref{sec:simulation}\\
                $\alpha$ & 0.995 & & Sect.~\ref{sec:simulation} and Eq.~\eqref{eq:cooling}\\
                $N_\mathrm{cycles}$ & 500 & & Sect.~\ref{sec:simulation}\\
                $N_\mathrm{moves}$ & 250\,000 & & Sect.~\ref{sec:simulation}\\
                \hline
        \end{tabular}
\end{table}

\begin{table*}
        \centering
        \caption{Summary of algorithm performance tests. We applied the tiling algorithm to the 4MOST mock catalogues. We ran the algorithm with the default parameters and with parameters where certain optimisation options were disabled. In each test, the same set of targets was expected to be completed with the same completion criteria. The table below gives the summary statistics for each generated tiling. As expected, the test with all optimisation enabled gives better results than tests with disabled optimisation options.}
        \label{tab:perform}
        \begin{tabular}{lccccccccc}
                \hline
                Test name & $N_\mathrm{tile}$ & $N_\mathrm{OB}$ & Mean $T_\mathrm{exp}$ & Mean $T_\mathrm{OB}$ & Sum of $T_\mathrm{exp}$ & Sum of $T_\mathrm{OB}$ & Obs. frac. & Extra $T_\mathrm{obs}$ & Extra $T_\mathrm{total}$\\
                 & & & min & min & hours & hours & \% & \% & \% \\
                 & (1) & (2) & (3) & (4) & (5) & (6) & (7) & (8) & (9) \\
                \hline
                Default$^a$      & 40503 & 12375 & 14.36 & 64.90 &
                              9694 & 13386 & 72.4 &  0.0 &  0.0 \\
                Fix\_PA$^b$      & 40729 & 12428 & 14.36 & 64.99 &
                              9751 & 13462 & 72.4 &  0.6 &  0.6 \\
                Fix\_Texp$^c$    & 35426 & 12283 & 17.70 & 67.24 &
                             10451 & 13765 & 75.9 &  7.8 &  2.8 \\
                Fix\_PA\_Texp$^d$ & 35920 & 12564 & 17.70 & 66.68 &
                             10596 & 13963 & 75.9 &  9.3 &  4.3 \\
                No\_OH$^e$       & 39305 & 19147 & 14.85 & 42.94 &
                              9731 & 13702 & 71.0 &  0.4 &  2.4 \\
                Not\_opt$^f$     & 36172 & 19909 & 17.70 & 43.65 &
                             10671 & 14485 & 73.7 & 10.1 &  8.2 \\
                \hline
        \multicolumn{10}{l}{$^a$ Default tiling with all optimisation options enabled.}\\
        \multicolumn{10}{l}{$^b$ Position angle of each tile is kept fixed during the MCMC run. Initial position angle for each tile is randomly determined.}\\
        \multicolumn{10}{l}{$^c$ Exposure time of each tile is fixed to 17.7~min, which allows three exposures during single OB.}\\
        \multicolumn{10}{l}{$^d$ Exposure time and position angle of tiles are kept fixed during the MCMC run.}\\
        \multicolumn{10}{l}{$^e$ Overhead fraction is not minimised during the optimal tiling generation.}\\
        \multicolumn{10}{l}{$^f$ Exposure time and position angle of tiles are kept fixed and overhead fraction is not minimised during the MCMC run.}\\
        \multicolumn{10}{l}{$(1)$ Number of tiles in the final tiling configuration after the MCMC run.}\\
        \multicolumn{10}{l}{$(2)$ Number of OBs in the final tiling configuration after the MCMC run.}\\
        \multicolumn{10}{l}{$(3)$ Mean exposure time of tiles in the final tiling configuration.}\\
        \multicolumn{10}{l}{$(4)$ Mean OB length (including overheads) in the final tiling configuration. Maximum OB length is 70~min.}\\
        \multicolumn{10}{l}{$(5)$ Sum of exposure times of all tiles in the final tiling configuration. Total observational time.}\\
        \multicolumn{10}{l}{$(6)$ Sum of exposure times and overheads associated with each tile and OB. Total telescope time with overheads.}\\
        \multicolumn{10}{l}{$(7)$ Fraction of total time that is spent for observations.}\\
        \multicolumn{10}{l}{$(8)$ Extra observational time (without overheads) that is needed compared with the Default tiling.}\\
        \multicolumn{10}{l}{$(9)$ Extra total telescope time (with overheads) that is needed compared with the Default tiling.}\\
        \end{tabular}
\end{table*}

To test the impact of $U_\mathrm{BGD}$ during the tiling pattern generation, we used all targets from the 4MOST mock catalogues. The distribution of the required exposure time in the sky is shown in Fig.~\ref{fig:texp_map}. The tiling pattern was generated to have roughly 50\,\% of dark tiles, 20\,\% of grey tiles and 30\,\% of bright tiles. The regions where these tiles should be located were not fixed beforehand. The tiling algorithm decides based on $T_\mathrm{exp}^\mathrm{BGD}(t)$ which sky regions should be observed during bright, grey or dark sky conditions. The fraction of tiles for each sky condition is a free parameter in the tiling algorithm and can be tuned as necessary. Table~\ref{tab:parameters} gives the parameters that were used during the test simulation.

Fig.~\ref{fig:tiling_fullsky} shows the output tiling pattern for the full sky. In the upper panel, the footprints of different sub-surveys in 4MOST are clearly visible, including the Milky Way in the middle of the image, the bulge region and the Magellanic Clouds. Regarding the division of tiles between predefined sky conditions, as expected, the Milky Way and the Magellanic Clouds are mostly observed during bright time because most of the stars are bright and the required exposure time per object is roughly the same for all sky conditions. Most of the extragalactic sky contains faint galaxies and AGNs, where the required exposure time depends strongly on the sky condition. For faint extended objects, the dark sky condition is highly preferred and the tiling algorithm assigns most of the dark time tiles to the extragalactic sky. Grey time is almost uniformly distributed and does not have any clear preference in Galactic or extragalactic sky. In general, the proposed tiling algorithm minimises the total time that is necessary to observe the given set of targets in the sky.

\subsection{Performance analysis of the tiling algorithm}

In this section we analyse how well the algorithm performs compared with slightly less optimised tiling solutions. In general, it is not straightforward to compare the proposed algorithm with other available methods. The main reason is that different algorithms optimise different aspects and it is not straightforward to define a common merit function (often called a ``metric''\footnote{Not to be confused with the differential geometry sense of ``metric'' that is fundamental to the spacetime of modern astronomy.}) that can be easily compared.

In Section~\ref{sec:mock} we presented two tiling solutions with and without an overhead minimisation (see Figs.~\ref{fig:sky_pixels} and~\ref{fig:sky_tiling}). While both of them required approximately the same amount of summed exposure time, the tiling solution without the overhead optimisation requires about 30\% more time for overheads. In this section we extend this analysis using the 4MOST mock catalogues and compare the proposed algorithm performance against itself.

We ran the algorithm several times. During each test run, we disabled one or many optimisation options. This allows us to estimate the effect of these optimisation options. During these tests, we either fixed the position angle of each tile, fixed the tiles exposure times, disabled the overhead minimisation or applied several of them together. Table~\ref{tab:perform} gives the summary statistics for these test runs. As expected, the tiling with all optimisation options enabled provides the best results. With some optimisation disabled, the final tiling configuration requires up to eight per cent more telescope time.

During all test runs we used the same 4MOST mock target catalogues and the generated tiling uses exactly the same completion criteria. Hence, all these test runs should provide roughly the same scientific outcome.\footnote{To estimate the real scientific merit of the generated tiling configurations requires full simulation of the 4MOST observations. The generated tiling pattern alone does not allow the estimation of the real scientific merit directly.} Although the generated tiling solutions are all slightly different, each one of these solutions constitutes an optimal solution given the parametrisation used in the optimisation process. The final tiling solution is mostly determined by the underlying target density. The disabled optimisation options have only a second order effect on the final solution. Using a naive tiling that does not follow the underlying target density would give a significantly worse solution.

To conclude, the optimal tiling solution is mostly driven by the underlying target density. The MCMC optimisation of the tile position angles, exposure times and minimisation of the overall overhead fraction gives up to an eight per cent improvement compared with the slightly less optimised tiling solutions.

\section{Conclusions and Discussion}
\label{sec:conclusions}

In this paper, we propose a tiling algorithm for multi-object spectroscopic surveys that is based on marked point processes. In the algorithm, the optimal tiling pattern is modelled as a marked point process where each tile is considered as a marked point or object. Finding the optimal tiling solutions is equivalent to finding the set of tiles with exposure times that is required to efficiently observe the targets given in the input catalogue. The optimisation problem is solved using a Metropolis-Hastings algorithm with simulated annealing.

The proposed algorithm finds an optimal tiling pattern given an input target catalogue. The algorithm finds the optimal tiling solution in the sky regions that are observed once or several times. Simultaneously, the algorithm finds an efficient solution in regions that should be visited multiple times. We found that the optimal tiling pattern selected by the algorithm follows the underlying target density very well. Hence, the algorithm can be used simultaneously for surveys that require multiple visits and for surveys that need uniform sky coverage.

The proposed algorithm does not assume a fixed exposure time per observation. Assuming that the required exposure time per target is available in the input catalogue data files, the algorithm determines a tentative exposure time for each tile, while taking the overhead time per each observation into account. In general, the algorithm allows to minimise the total time that is needed to successfully and efficiently observe the objects given in the input target catalogue. Additionally, the algorithm can divide the tiles between different sky conditions, assuming that the exposure time per target as a function of sky condition is available.

Finding an optimal tiling solution requires a clear definition of a merit function that should be maximised. In the proposed algorithm, the merit function is defined via an energy function, where the energy function takes different aspects of the optimal tiling problem into account. The energy function defined in this paper optimises the fibre-to-target assignment, minimises the total overhead time, includes interactions between tiles, and forces the tiles to be divided between predefined sky conditions. The balance between these components can be fine tuned in the algorithm, based on the input catalogue and the survey science goals.

The proposed algorithm is tested using the current mock catalogues of the 4MOST consortium surveys, covering the Milky Way and extragalactic sky. We show that the generated optimal tiling pattern matches the estimated required exposure time as a function of sky coordinates very well. The optimal tiling pattern follows the edges of different sub-survey patches in the sky, allowing the generation of an efficient tiling that takes the target density variations in the sky naturally into account. The generated tiling pattern is used together with the probabilistic fibre-to-target assignment algorithm proposed in \citet{Tempel:2019}, showing very high fibre-usage efficiency and survey completeness. In general, the optimal tiling algorithm proposed in this paper is an input for the probabilistic fibre-to-target assignment algorithm described in \citet{Tempel:2019}.

The marked point process framework behind the proposed tiling algorithm is very flexible and allows the redefinition of the described energy function components or the introduction of new components. For example, the interaction between tiles can be used to minimise the gaps between individual tiles and to construct a tiling pattern that uniformly covers a contiguous area of sky. When necessary, the interaction energy can also be used to force gaps between tiles in order to cover larger sky areas with the same number of tiles. Depending on the survey science case, an appropriate interaction energy for an optimal tiling pattern can be chosen. The exact definition of the interaction energy for the 4MOST survey will be determined during the 4MOST survey optimisation phase.

The proposed algorithm generates a tiling pattern that is needed to most efficiently observe the given set of targets in the input catalogue. However, the tiling algorithm does not determine when each tile should be observed. Neither does it constrain how much time is available for the 4MOST survey. To solve this problem, one needs a scheduling algorithm that determines which tiles should be observed and when they should be observed. The scheduling problem can be solved independently of the tiling challenge. A good scheduling algorithm for the 4MOST survey will be developed during the 4MOST survey preparation and is not part of the algorithm proposed in this paper.

Regarding the division of tiles between various predefined sky conditions, in the algorithm the fraction of time that is available during dark, grey or bright sky conditions is currently considered. In reality, the division between predefined sky conditions should also take the distribution of tiles in the sky into account. This is necessary, since certain sky regions are only visible during the summer or winter periods and the algorithm should generate tiles with various sky conditions everywhere in the sky. This shows one possible improvement for the proposed tiling algorithm that still needs to be studied. For the 4MOST survey, the need for this improvement will be assessed in combination with the scheduling algorithm. A simple solution is to observe some tiles during better sky conditions than those assigned by the algorithm and to scale the exposure time per tile accordingly. A more optimal but time-consuming solution is to fine-tune the tiling algorithm parameters so that the produced distribution of tiles with predefined sky conditions follows the fraction of available time in different sky regions.

Table~\ref{tab:parameters} gives the free parameters of the tiling algorithm that should be determined for an optimal tiling solution. Many of these parameters affect the speed and convergence of the algorithm and have only a minor impact on the final tiling solution. However, some of the parameters have a direct impact on the optimal tiling solution and should be determined while taking the input target catalogue and survey science goals into account. One example is the parameters $c_\mathrm{LR}$ and $c_\mathrm{HR}$ that determine the importance between the numbers of low-resolution and high-resolution targets. In this paper, constant values are assumed across the sky. However, if a sky region is dominated by LR targets, then the optimisation should take that into account. This can be achieved by defining different $c_\mathrm{LR}$ and $c_\mathrm{HR}$ values in different parts of the sky. These optimisations depend very strongly on the input target catalogue and will be included in the algorithm during the 4MOST survey preparation.

Computationally, the proposed algorithm is somewhat demanding. In the 4MOST survey, we have approximately 40\,000 individual tiles and Markov-chain Monte Carlo (MCMC) sampling and the optimisation of large number of tiles take some time. Additionally, during each MCMC step we have to perform the fibre-to-target assignment. In the proposed algorithm we use a statistical fibre-to-target assignment, which helps to improve the speed of the algorithm significantly. Despite that, to find an optimal tiling pattern for the full 4MOST survey (about 50 million targets and 40 thousand tiles) takes currently up to a few days using 24 cores on a shared memory machine. The algorithm scales reasonably well using OpenMP parallelisation. It is not yet tested, how well the algorithm scales using MPI parallelisation. The computation time can be potentially reduced by using better optimisation and parallelisation. During real observations, the tiling algorithm should be run at the beginning of the survey, in which case a few days of computational time is not a problem. However, during the execution of the survey, one might want to rerun the tiling algorithm to optimise the tiling that better matches the remaining targets, or because the input target catalogue has been updated. In these cases, one does not have to run the tiling algorithm from scratch. The MCMC sampling of the tiling pattern can be initialised using the previous tiling solution. This will significantly reduce the computational cost and allow the tiling pattern to be updated during the survey within a reasonable amount of computational time.

To conclude, the tiling algorithm presented in this paper is a new approach to solving the optimal tiling challenge for multi-object spectroscopic surveys. The current algorithm is a proposed solution for the 4MOST survey and in combination with the probabilistic fibre-to-target assignment presented by \citet{Tempel:2019} solves two major challenges faced during the 4MOST survey preparation. With appropriate modifications, the algorithm that we propose can be also applied to other forthcoming multi-object spectroscopic surveys.

\section*{Acknowledgements}

This work has made use of the development effort for 4MOST, an instrument under construction by the 4MOST Consortium (\url{https://www.4most.eu/cms/consortium/}) for the European Southern Observatory (ESO).
Part of this work was supported by institutional research funding IUT40-2 of the Estonian Ministry of Education and Research. We acknowledge the support by the Centre of Excellence ``Dark side of the Universe'' (TK133) and by the grant MOBTP86 financed by the European Union through the European Regional Development Fund. Part of this work was supported by the ``A next-generation worldwide quantum sensor network with optical atomic clocks'' project, which is carried out within the TEAM IV programme of the Foundation for Polish Science co-financed by the European Union under the European Regional Development Fund. This work has been supported by the Polish MNiSW grant DIR/WK/2018/12. AK and NC acknowledge funding by the DFG -- Project-ID 138713538 -- SFB 881 (``The Milky Way System''), subprojects A03, A05, A09, A11. MRC acknowledges funding from the European Research Council (ERC) under the European Union's Horizon 2020 research and innovation programme (grant agreement no. 682115). GT was supported by the grant The New Milky Way from the Knut and Alice Wallenberg foundation, by the grant 2016-03412 from the Swedish Research Council, and by the Swedish strategic research programme eSSENCE.

\section*{Data availability}

The 4MOST mock catalogues used in this article are subject to the data access policies of the 4MOST consortium. The software code will be shared on reasonable request to the corresponding author.


\bibliographystyle{mnras_openaccess}
\bibliography{mybib}


\bsp    
\label{lastpage}
\end{document}